\documentclass[aps,nofootinbib,superscriptaddress,twocolumn,prc]{revtex4-1}

\usepackage{hyperref,amsmath,graphicx,url,cancel}
\usepackage[latin1]{inputenc}
\usepackage{bm}
\usepackage[normalem]{ulem}

\newcommand{\bfrac}[2]{\left(\frac{#1}{#2}\right)}

\newcommand{\infrac}[2]{{#1}/{#2}}
\newcommand{\inbfrac}[2]{\left({#1}/{#2}\right)}
\newcommand{\derr}[2]{\frac{d{#1}}{d{#2}}}
\newcommand{\inderr}[2]{d{#1}/d{#2}}
\newcommand{\inbderr}[2]{\left(d{#1}/d{#2}\right)}
\newcommand{\parr}[2]{\frac{\partial{#1}}{\partial{#2}}}
\newcommand{\bparr}[2]{\left(\frac{\partial{#1}}{\partial{#2}}\right)}
\newcommand{\inparr}[2]{\partial{#1}/\partial{#2}}
\newcommand{\inbparr}[2]{(\partial{#1}/\partial{#2})}

\newcommand{\txt}[1]{\textrm{#1}}

\usepackage{xcolor}

\begin{document}

\title{Sensitivity of Au+Au collisions to the symmetric nuclear matter equation of state \\at 2--5 nuclear saturation densities}

\author{Dmytro Oliinychenko}
\email{dmytrooliinychenko@gmail.com}
\affiliation{Institute for Nuclear Theory, University of Washington, Box 351550, Seattle, Washington 98195, USA}

\author{Agnieszka Sorensen}
\email{agnieszka.sorensen@gmail.com}
\affiliation{Institute for Nuclear Theory, University of Washington, Box 351550, Seattle, Washington 98195, USA}

\author{Volker Koch}
\affiliation{Lawrence Berkeley National Laboratory, 1 Cyclotron Road, Berkeley, California 94720, USA}

\author{Larry McLerran}
\affiliation{Institute for Nuclear Theory, University of Washington, Box 351550, Seattle, Washington 98195, USA}

\begin{abstract}
We demonstrate that proton and pion flow measurements in heavy-ion collisions at incident energies ranging from 1 to 20 GeV per nucleon in the fixed target frame can be used for an accurate determination of the symmetric nuclear matter equation of state  at baryon densities equal 2--4 times nuclear saturation density $n_0$. We simulate Au+Au collisions at these energies using a hadronic transport model with an adjustable vector mean-field potential dependent on baryon density $n_B$. We show that the mean field can be parametrized to reproduce a given density dependence of the speed of sound at zero temperature $c_s^2(n_B, T = 0)$, which we vary independently in multiple density intervals to probe the differential sensitivity of heavy-ion observables to the equation of state at these specific densities. 
Recent flow data from the STAR experiment at the center-of-mass energies $\sqrt{s_{\txt{NN}}} = \{3.0, 4.5 \}\ \rm{GeV}$ can be described by our model, and a Bayesian analysis of these data indicates a hard equation of state at $n_B \in (2,3) n_0$ and a possible phase transition at $n_B \in (3,4) n_0$. 
More data at $\sqrt{s_{\txt{NN}}} = 2$--$5~ \txt{GeV}$, as well as a more thorough analysis of the model systematic uncertainties will be necessary for a more precise conclusion.
\end{abstract}

\maketitle

\section{Introduction}

The baryon density is approximately $n_B \approx 0.16 ~ \txt{fm}^{-3}$ at the center of a nucleus and averages to about $0.12 ~\txt{fm}^{-3}$ over its entire volume, almost independently of the size of the given nucleus \cite{WongIntroductoryNuclearPhysics}. Neglecting the finite size effects in nuclei as well as Coulomb interactions, one arrives at an idealized theoretical concept of \textit{nuclear matter}, which is in equilibrium at density $n_0 \approx 0.16$ fm$^{-3}$, often called the nuclear matter saturation density \cite{WongIntroductoryNuclearPhysics, Bethe:1971xm}. The only way of obtaining a substantially more dense nuclear matter in a laboratory is to collide heavy nuclei at relativistic incident velocities. Such collisions of two nuclei result in a rapid (timescales on the order of a few fm/$c$) compression and heating, followed by expansion and cooling of the produced fireball. The outcomes of both the compression and the expansion phase depend on the strong interactions between the constituents of the fireball. In equilibrium, the influence of these interactions on the properties of the medium is described by the equation of state (EOS), that is by the dependence of the equilibrium pressure $P$ on temperature $T$, net baryon density $n_B$, net strangeness density $n_S$, and net charge density $n_Q$\footnote{Depending on the problem at hand, it is sometimes more convenient to represent the EOS in the form $P(T, \mu_B, \mu_S, \mu_Q)$, where $\mu_i$ denotes the chemical potential associated with the conserved charge $n_i$ and $i \in \{B, S, Q \}$; in the form $P(\mathcal{E}, n_B, n_S, n_Q)$, where $\mathcal{E}$ is the energy density; or in the form $P(\mathcal{E}, n_B, n_S, n_{I3})$, where instead of the charge density $n_Q$ one considers the isospin projection density $n_{I3}$, with the two densities closely related through the relation between the electric charge and the isospin projection of a hadron, $Q = I_3 + \frac{1}{2}(B+S)$.}.

We note that while the EOS is only defined in equilibrium, the systems created in heavy-ion collisions may well be out-of-equilibrium. To incorporate non-equilibrium effects in simulations, one can either introduce viscous corrections to a fluid-dynamic description, which works well for small deviations from equilibrium, or one can resort to transport theory, which accounts for the full non-equilibrium dynamics governed by the Boltzmann equation. Within transport-theoretical approaches, the model parameters (in particular, the mean-field interaction) can be adjusted to reproduce relevant observables measured in heavy-ion reactions such as flow. The EOS is then obtained by using these model parameters in the transport equation, which under the assumption that the system is in equilibrium can be used to calculate, e.g., the pressure.

\begin{figure}[tbh]
    \centering
    \includegraphics[width=0.49\textwidth]{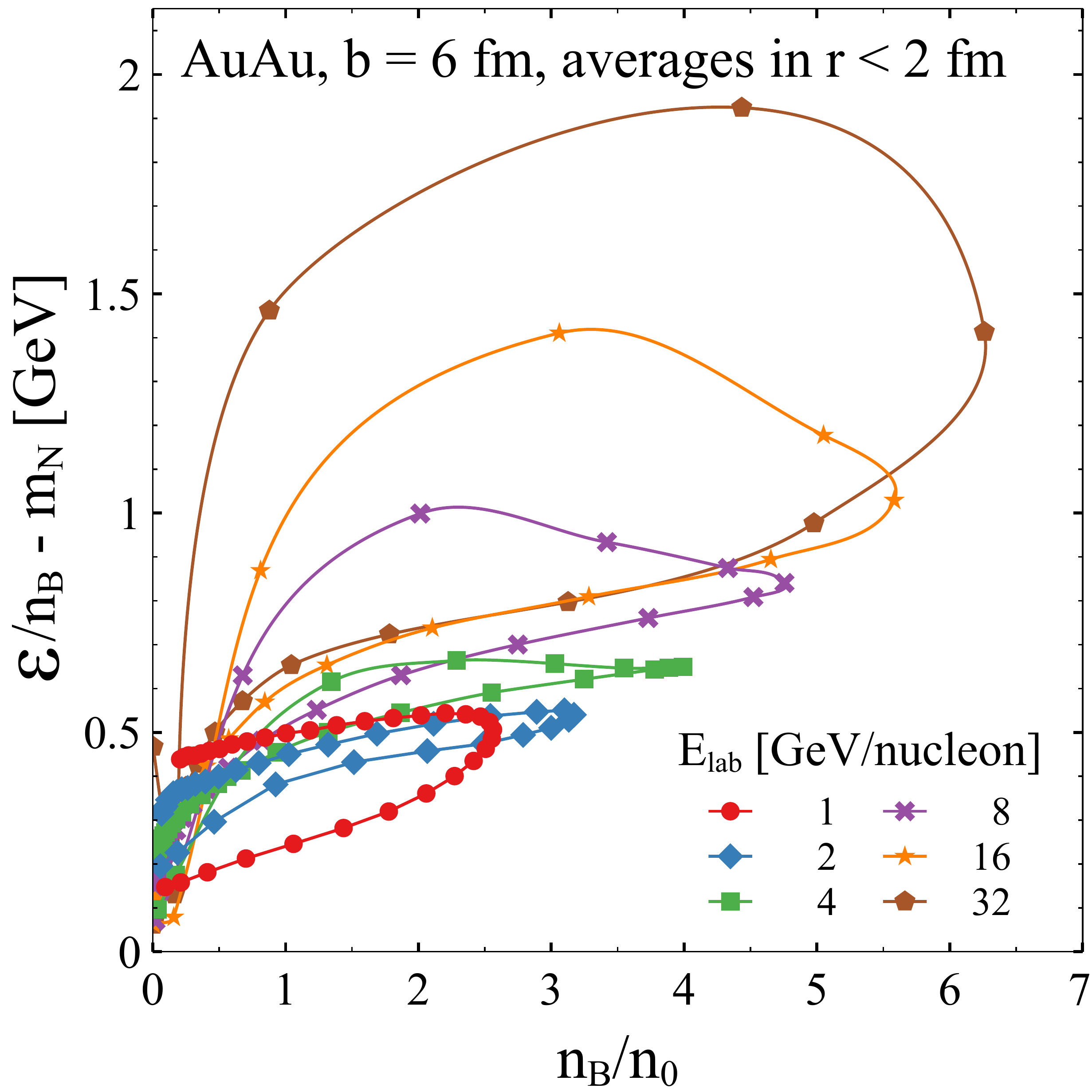}
    \caption{Trajectories of a central region (i.e., a sphere of radius $r = 2~ \txt{fm}$ located at the event-averaged center-of-mass of the collision system) in Au+Au collisions at impact parameter $b = 6~\rm{fm}$, obtained from simulations in \texttt{SMASH}. The contours demonstrate qualitatively which ranges of densities are probed at which collision energies. The simulations are performed in the center-of-mass frame and the time $t$ is also given in that frame; the time difference between the shown data points is 1 fm/$c$, and the first points shown are for $t = -2$ fm/$c$ when nuclei are not yet touching (by the convention used in \texttt{SMASH}, $t = 0$ is the time at which nuclei would touch in a central collision; note that for mid-central collisions, the center of mass of the system at $t=0$ is between the nuclei, which explains the corresponding low values of density). For collisions at $E_{\txt{lab}} = 1$ and 2 GeV/nucleon (red and blue marks, respectively) the trajectories are traversed in the counterclockwise direction with increasing time, while for collisions at 4 GeV/nucleon and above, the trajectories are traversed in the clockwise direction.  The energy density $\mathcal{E}$ is calculated from the $00$-component of the energy-momentum tensor $T^{\mu\nu}$ in the Landau frame; note that mean-field contributions are not taken into account in this calculation of $T^{\mu\nu}$, that is $\mathcal{E} = \mathcal{E}_{\txt{kin}}$, to enable a clearer inference of the different temperatures reached in collisions at different energies. The simulations used a mean-field potential parametrized to reproduce the standard Skyrme EOS at $n_B \in [0,2]n_0$ and requiring that $c_s^2(n_B, T = 0) = 0.3$ at higher densities (see Sec.~\ref{sec:parametrization} for details of the EOS construction). Changes in the trajectories due to employing different EOSs are shown in Fig.\ \ref{fig:II}. }
    \label{fig:I}
\end{figure}

Ultimately, one of the major goals of heavy-ion collision experiments is to extract the EOS within the experimentally accessible domain. This domain is, admittedly, limited; for example, $n_S \approx 0$ in heavy-ion collisions. Even more importantly, compression is always accompanied by heating, and as a result regions characterized by high $n_B$ and low $T$ are not probed at any collision energy \cite{Arsene:2006vf}. This can be seen in Fig.\ \ref{fig:I}, which shows phase trajectories of the central region of a heavy-ion collision at different values of the incident kinetic energy per nucleon (excluding the rest mass) $E_{\txt{lab}}$, obtained from simulations using the hadronic transport code \texttt{SMASH} \cite{Weil:2016zrk} (version 2.1 \cite{smash_version_2.1}).

\begin{figure}[tbh]
    \centering
    \includegraphics[width=0.49\textwidth]{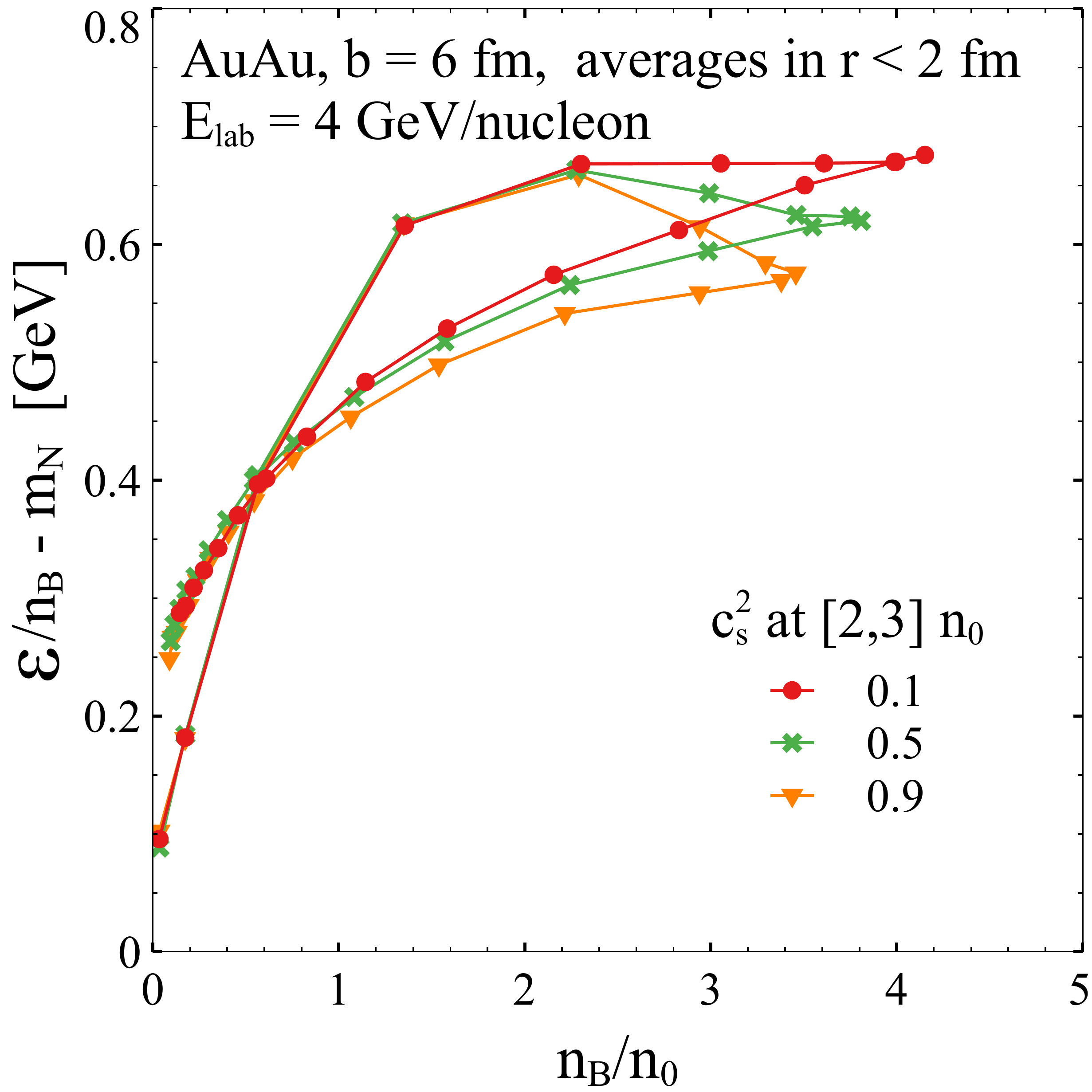}
    \caption{Phase trajectories in a Au+Au collisions at $E_{\txt{lab}} = 4~ \txt{GeV/nucleon}$, obtained using the same simulation setup as in Fig.\ \ref{fig:I}, demonstrating that varying the EOS only in the density region $n_B \in (2, 3)n_0$ leads to a higher (lower) reached values of the density and energy per nucleon in the case of the softer (harder) EOS. Here, in contrast to previous works where the stiffness of the EOS is controlled solely by the value of nuclear matter incompressibility at saturation $K_0$ (see, e.g., \cite{Danielewicz:2002pu,Fuchs:2000kp}), the stiffness of the EOS is changed by varying the value of the speed of sound $c_s^2$ at $T = 0$ for $n_B \in (2, 3)n_0$, with small (large) values of $c_s^2$ corresponding to a relatively soft (hard) EOS in that density range. The trajectories are traversed in the clockwise direction with increasing time, and the time difference between the shown data points is 1 fm/$c$. Below $n_B = 2n_0$, the mean-field potential is parametrized to reproduce the standard Skyrme EOS, while above $n_B = 2n_0$, the mean-field is parametrized to reproduce values of $c_s^2$ from a representative set, $c_s^2 (n_B > 2n_0, T = 0) = \{0.1, ~0.5, ~0.9\}$ (see Sec.~\ref{sec:meth} for details of the EOS construction). }
    \label{fig:II}
\end{figure}

The role of the EOS in heavy-ion collisions is intuitively understandable: a stiffer EOS (meaning an EOS with a relatively large $\inparr{P}{n_B}$ or $\inparr{P}{\mathcal{E}}$) leads to less compression, less heating, and a faster transverse expansion than a softer EOS (see, e.g., \cite{Fuchs:2000kp}). In Fig.\ \ref{fig:II}, we not only demonstrate that this is indeed the case in our simulations, but we also show that there is a large sensitivity to changing the stiffness of the EOS only in a specific density region (we will elaborate on this point in the following sections). In terms of observables, the slower expansion characterizing a softer EOS is expected to produce lower mean transverse momenta (see, e.g., \cite{Steinheimer:2022gqb}), while the increased heating in this case leads one to expect a larger thermal dilepton and photon yield (see, e.g., \cite{Huovinen:1998tq,Savchuk:2022aev,Li:2022iil}). Moreover, as the fireball lifetime is longer for a softer EoS, one would also expect that the combination of the femtoscopic radii $R_{\txt{out}}^2 - R_{\txt{side}}^2 $, which can be shown to be proportional to the duration of the emission of detected particles \cite{Bertsch:1989vn,Pratt:1990zq}, will be larger in this case. 

Nevertheless, the most stringent currently available constraints on the symmetric nuclear matter EOS come from angular distributions in the transverse plane $\infrac{dN}{d\phi}$, where $\phi$ denotes the azimuthal angle, which are highly sensitive to the EOS and, at the same time, measurable with high precision \cite{Danielewicz:1985hn}. Indeed, the sensitivity of the flow observables has been demonstrated by multiple hydrodynamic \cite{Stoecker:1980vf,Ollitrault:1992bk,Rischke:1995pe,Stoecker:2004qu,Brachmann:1999xt,Csernai:1999nf,Ivanov:2014ioa} and hadronic transport \cite{Hartnack:1994ce,Li:1998ze,Danielewicz:2002pu,LeFevre:2015paj,Wang:2018hsw,Colonna:2020euy,Nara:2021fuu} models. We notice, however, that most of these works compare only several EOSs (often just an EOS with a phase transition to a quark-gluon plasma to an EOS without such transition), and do not attempt to quantify the sensitivity by parametrizing the EOS continuously and constraining the parameters. The works that do explore a range of possible EOSs \cite{LeFevre:2015paj,Wang:2018hsw,Danielewicz:2002pu} parametrize the EOS with a single parameter: the incompressibility of isospin-symmetric nuclear matter at $n_B = n_0$ and $T=0$, defined as $K_0 \equiv 9\inbfrac{\partial P}{\partial n_B}|_{n_B = n_0}$ or equivalently as $K_0 = 9 n_B^2 \inbfrac{\partial^2 (E/A) }{\partial n_B^2}|_{n_B = n_0}$, where $E/A$ is energy per baryon. In this work, we aim at exploring the sensitivity of the flow observables to the EOS parametrized by varying both the incompressibility $K_0$ and values of the speed of sound at different ranges of baryon density: $c^2_{[2,3]n_0} \equiv c_s^2 \big[n_B \in (2,3)n_0\big]$, $c^2_{[3,4]n_0} \equiv c_s^2 \big[n_B \in (3,4)n_0\big]$, and $c^2_{[4,5]n_0} \equiv c_s^2 \big[n_B \in (4,5)n_0\big]$. We consider this work as the first step towards a full Bayesian analysis of all available observables with a flexible and suitably parametrized EOS. Here, however, we use only a limited set of measurements, that is flow measurements as provided by the E895 Collaboration \cite{E895:1999ldn,E895:2000maf,E895:2001axb,E895:2001zms} and by recent results from the STAR Collaboration \cite{STAR:2020dav,STAR:2021yiu} based on Phase II of the Beam Energy Scan (BES) program at the Relativistic Heavy Ion Collider (RHIC).

The effects of the EOS on $\infrac{dN}{d\phi}$ are known to be substantial in noncentral collisions at energies for which the speed of the spectators is comparable to that of the fireball expansion \cite{Sorge:1996pc,Danielewicz:2002pu}. The $\infrac{dN}{d\phi}$ distribution of protons around midrapidity $y' = 0$ (where $y'=y/y_{\txt{beam}}$ is the center-of-mass rapidity $y$ scaled by the beam rapidity in the center-of-mass frame 
\footnote{
Rapidity is defined as $y (\beta) = \frac{1}{2} \ln \big[\infrac{(1+\beta)}{(1-\beta)}\big]$, where $\beta$ is the velocity, and a Lorentz boost corresponding to the relative velocity of two inertial frames $\beta_0$ is equivalent to a shift of $y$ by $y (\beta_0)$. Therefore, the center-of-mass rapidity $y$ for collisions in the fixed-target mode can be obtained by taking $y \to y - y_{\txt{beam}}$. In the center-of-mass frame, rapidity distributions of measured particles are centered around $y=0$, while the beam rapidities are at $\pm y_{\txt{beam}}$. To make results at different collision energies, and therefore different spreads in rapidity, easier to compare, one uses the scaled rapidity $y' = y/y_{\txt{beam}}$, since for the colliding beams one then always has $y'_{\txt{beam}} = \pm 1$ in the center-of-mass frame. Notice that this property of $y'$ is satisfied only in the center-of-mass frame.
}) has maxima coincident with the reaction plane in the case where the spectators move out of the way of the expanding fireball fast enough. In the opposite case, where the spectators block the in-plane fireball expansion, the preferential emission occurs in the out-of-plane direction (due to the role of the spectators, this phenomenon is often called a ``squeeze-out''). This behavior is captured by the second Fourier coefficient of $\infrac{dN}{d\phi}$,
\begin{eqnarray}
v_2 = \langle \cos 2\phi \rangle = \frac{\int d\phi ~ \cos (2\phi) ~\frac{dN}{d\phi}  }{ \int d\phi ~  \frac{dN}{d\phi} } ~,
\end{eqnarray}
known as the elliptic flow, which quantifies the difference between the in-plane and out-of-plane emission. A positive $v_2$ indicates a preferential emission toward angles $ \phi \approx 0$ and $\phi \approx \pi$, while a negative $v_2$ indicates a preferential emission toward $\phi \approx \pi/2$ and $\phi \approx 3\pi/2$. For collisions at $E_{\txt{lab}} \approx 1$--$10 ~ \txt{GeV/nucleon}$, for which the spectators still occupy the vicinity of the collision region during the expansion phase, a faster fireball expansion due to a stiffer EOS will be correspondingly more forcefully blocked by the spectators, resulting in a larger squeeze-out effect and a more negative $v_2$. 

Another observable used to constrain the EOS is the slope of the directed flow at mid-rapidity, 
\begin{eqnarray}
\frac{dv_1}{dy'}\bigg|_{y'=0} = \frac{d\langle \cos\phi \rangle}{ dy'} \bigg|_{y'=0}~,
\end{eqnarray}
where the directed flow $v_1$ is defined as the first Fourier coefficient of $\infrac{dN}{d\phi}$, so that a positive $v_1 = \langle \cos\phi \rangle$ characterizes a preferential emission in the $\phi \in [-\pi/2,\pi/2]$ direction compared to the $\phi \in [\pi/2, 3\pi/2]$ direction. Intuitively, $dv_1/dy'$ quantifies the strength of spectator deflection \cite{Voloshin:2008dg}. To visualize the physical meaning of $dv_1/dy'$, one can imagine that spectators are two masses pushed apart by a spring oriented along the transverse plane, where the spring represents the compressed fireball. In the center-of-mass frame, the spectators moving in the positive $y$ direction will be deflected towards $\phi \in [-\pi/2, \pi/2]$, while the spectators moving in the negative $y$ direction will be deflected towards $\phi \in [\pi/2, 3\pi/2]$, resulting in a positive $v_1$ for $y >0$ and a negative $v_1$ for $y<0$ \footnote{Note that this picture has a certain dependence on how one defines the coordinate axes in the problem. Here, we define the impact parameter vector $\bm{b}$ (considered in the center-of-mass frame, at the moment of the closest approach of the centers of the two nuclei) as starting at the center of the heavy ion moving in the negative $y$ direction and ending at the center of the heavy-ion moving in the positive $y$ direction, and we define the $x$ axis of the transverse plane to be parallel to the so-defined $\bm{b}$. The azimuthal angle $\phi$ is then measured from the $x$ axis. A realization of this system of coordinates, in which the cross product of the basis vectors of the transverse plane is parallel to the $z$ axis, is depicted in Fig.\ 1 of \cite{Danielewicz:2002pu}. In general, by convention, one defines the coordinate axes such that $dv_1/dy$ is positive for spectators deflected away from the collision zone.}. As can be easily visualized using the spring analogy, a softer EOS results in a weaker deflection and a smaller $dv_1/dy'$. Moreover, it has been shown, both in hydrodynamic and transport simulations, that a sufficiently soft EOS (including EOSs with phase transitions) can lead to a negative $dv_1/dy'$ \cite{Stoecker:2004qu,Zhang:2018wlk}, which within our analogy corresponds to a situation in which the spring pulls the two masses closer together.

One of the most prominent constraints on the EOS using the flow analysis of heavy-ion collisions \cite{Danielewicz:2002pu} is based on comparing a hadronic transport model \cite{Danielewicz:1991dh,Danielewicz:1999zn} and flow data measured in the E895 experiment \cite{E895:1999ldn,E895:2000maf}. The main result of the work \cite{Danielewicz:2002pu} is constraining the incompressibility $K_0$ of symmetric nuclear matter, the only parameter of the EOS varied in that study, to be between 210 and 380 MeV. (We note here that in studies applying such simple parametrizations of the EOS to heavy-ion collisions at relativistic energies, which primarily probe large baryon densities, the incompressibility is treated as a parameter which specifies the behavior of the EOS at densities above the saturation density $n_0$.) Such a spread in the extracted values of $K_0$ originates not from data uncertainties, but from the fact that the used model was not able to describe the $v_2$ and $dv_1/dy'$ data simultaneously. While reproducing the $v_2$ measurements required larger values of $K_0$ (and therefore a harder EOS), the $dv_1/dy'$ data required smaller values of $K_0$ (corresponding to a softer EOS). In this work, we test whether a relaxation of the EOS model that allows one to independently vary the stiffness of the EOS in different density regions can lead to a more consistent description of the flow data and, consequently, to a more precise constraint on the EOS. For example, if densities to which $dv_1/dy'$ is most sensitive are higher than the corresponding densities for $v_2$, then we may make the EOS soft at high densities to reproduce the measured $dv_1/dy'$, and stiff at lower densities to reproduce the measured $v_2$. Such idea was, in fact, suggested in \cite{Danielewicz:2002pu}, but the chosen parametrization of the EOS did not allow to implement it. In this work, we construct a more flexible EOS parametrization and use it to fit the E895 data available at the time of \cite{Danielewicz:2002pu} as well as the newest STAR data for the same energy range \cite{STAR:2020dav,STAR:2021yiu}.

The structure of this work is as follows: Section \ref{sec:meth} gives a short overview of the dynamical evolution in hadronic transport (Sec.~\ref{sec:simulation_framework}), introduces a parametrization of the mean-field potential reproducing a given behavior of the speed of sound as a function of baryon density (Sec.~\ref{sec:parametrization}), and provides a summary of simulation and implementation choices relevant to the study at hand (Sec.~\ref{sec:details_of_the_simulations}). Section~\ref{sec:sensitivity} uses the developed model to explore the sensitivity of the flow observables to the stiffness of the EOS in separate density intervals. Section~\ref{sec:E895_and_STAR} presents our exploration of describing the available flow data from E895 and STAR experiments, where we conclude that there is a significant discrepancy between the results from the two data sets. Finally, Sec.~\ref{sec:only_STAR} discusses the results of our Bayesian analysis of the STAR flow data, including a discussion of the slight tension with the EOS inferences from neutron star observations. We summarize and provide an outlook in Section \ref{sec:summary}.

\section{Methodology}
\label{sec:meth}

In the following we use natural units in which $\hbar = c = k_B = 1$.

\subsection{Simulation framework}
\label{sec:simulation_framework}

As already discussed above, at projectile kinetic energies $E_{\txt{lab}}$ below 20 GeV per nucleon in the fixed target frame, corresponding to $\sqrt{s}_{\txt{NN}} \lesssim 6.4~\txt{GeV}$, spectators play a very important role both for $dv_1/dy'$ and for $v_2(y'=0)$, which was also explicitly demonstrated in hadronic transport simulations \cite{Zhang:2018wlk}. This is in contrast to the highest RHIC ($\sqrt{s_{\txt{NN}}}$ = 200 GeV) and LHC ($\sqrt{s_{\txt{NN}}}$ = 2.76 and 5.02 TeV) energies, where mid-rapidity observables are unaffected by the spectators. In consequence, most of the state-of-the-art hydrodynamic codes, intended for very high energies and neglecting spectators, are not applicable at $E_{\txt{lab}} = 2$--$20$ GeV/nucleon without modifications. Therefore, we choose to employ a hadronic transport simulation in which spectators are naturally included throughout the evolution. Our particular code of choice is the transport code \texttt{SMASH} \cite{Weil:2016zrk} (version 2.1 \cite{smash_version_2.1}), modified according to the prescription given in the next subsection. \texttt{SMASH} is a relativistic Boltzmann--Ueling--Uhlenbeck (BUU) type of hadronic transport with vector-density dependent mean-field potentials, which means that it is a Monte Carlo solver of the following kinetic equations
\begin{equation} 
\Pi_{\mu} \partial^{\mu}_{x} f_i(x,p) + \Pi_{\nu} \big(\partial_{\mu}^x A^{\nu}\big) \partial_p^{\mu} f_i(x,p) = I_{\txt{coll}}^{(i)} \,.
\label{eq:I}
\end{equation}
Here, $x^{\mu}$ and $p^{\mu}$ are position and momentum four-vectors (with indices suppressed where it is beneficial for clarity), $\Pi^{\mu}(x,p) \equiv p^{\mu} - A^{\mu}(x)$ is the kinetic momentum and $A^{\nu}$ is a vector field dependent on the baryon current as
\begin{eqnarray} \label{eq:vector_field}
A^{\mu} = \alpha(n_B) j_B^{\mu}\,,
\end{eqnarray}
where $\alpha(n_B)$ is a chosen function of the rest frame density $n_B \equiv \sqrt{(j_B)_{\mu}(j_B)^{\mu}}$, $f_i$ is the distribution function for the particle species $i$, and $I_{\txt{coll}}^{(i)}$ is the collision integral for the $i$th particle species. The distribution function $f_i$ depends on position as well as energy and momentum, and can be written in a form that accounts for the mass-shell condition explicitly:
\begin{eqnarray} 
f_i(x^{\mu}, p^{\mu})  = 2(2\pi) \Theta(\Pi^0) \delta\big(\Pi^{\mu} \Pi_{\mu} - m_i^2\big) ~
\tilde{f}_i(x^{\mu}, \bm{p}) ~.
\label{eq:III}
\end{eqnarray}
In the simulation, the continuous distribution function $\tilde{f}_i(x, \bm{p})$ describing a system of $N_i$ particles of species $i$ is approximated by using a standard test-particle ansatz, that is by taking $\tilde{f}(x,\bm{p})$ to be of the form
\begin{eqnarray} 
\hspace{-2mm}\tilde{f}_i(x, \bm{p}) = \frac{1}{N_T} \sum_{j=1}^{N_T N_i} \delta^{(3)}\big(\bm{x} - \bm{x}_j(t)\big) \delta^{(3)}\big(\bm{p} - \bm{p}_j(t)\big) ~.
\label{eq:IIIb}
\end{eqnarray}
The above equation means that the continuous distribution function $\tilde{f}_i(x, \bm{p})$, describing a physical system composed of $N_i$ particles, is sampled by $N = N_iN_T$ discrete points, or test particles, where the factor $N_T \gg 1$ is known as the number of test particles per particle or the oversampling factor. Intuitively, the larger the number of samples (i.e., the larger $N_T$ is used) is, the better is the description of $\tilde{f}_i(x, \bm{p})$. To preserve the essential properties of a system of $N_i$ particles in a simulation involving $N_T N_i$ test particles (such as the collision rate per test particle or the local density), all collision cross sections are reduced by a factor of $N_T$, and a single test-particle contributes to the energy density and charge densities with a factor of $1/N_T$.

By substituting Eqs.\ \eqref{eq:III} and \eqref{eq:IIIb} into the Vlasov equation (that is, the Boltzmann equation, Eq.\ \eqref{eq:I}, with the collision term set to zero, $I_{\txt{coll}}^{(i)}$ = 0), one obtains the following equations of motion for a single test particle in a system described by $f_i(x,p)$:
\begin{align} 
\frac{dx^{\mu}}{dt} &= \frac{\Pi^{\mu}}{\Pi^0} \label{eq:EOM1} ~, \\
\frac{d\Pi^{\mu}}{dt} &= \frac{\Pi_{\nu}}{\Pi^0} F^{\mu\nu} ~, \label{eq:EOM2} 
\end{align}
where $F^{\mu\nu}  = \partial^{\mu}A^{\nu} - \partial^{\nu}A^{\mu}$ and each test particle satisfies the mass-shell condition $\Pi^{\mu} \Pi_{\mu} = m_i^2$; details of this derivation can be found in Appendix \ref{app:test-particle_ansatz}. Propagating the test particles according to equations of motion, Eqs.\ \eqref{eq:EOM1}-\eqref{eq:EOM2}, together with performing decays and particle-particle collisions effectively solves Eq.\ \eqref{eq:I}, as at each time $t$ the distribution function is given by \eqref{eq:IIIb}. Importantly, while Eqs.\ \eqref{eq:EOM1}-\eqref{eq:EOM2} have the same form as standard relativistic equations of motion in an external vector field, in our case the vector field is directly dependent on local baryon density, that is, the equations governing the system are self-consistent. 

An alternative derivation of Eqs.\ \eqref{eq:EOM1}-\eqref{eq:EOM2} within a relativistic vector density functional (VDF) model of dense nuclear matter is shown in \cite{Sorensen:2020ygf}, where the equilibrium thermodynamics and transport equations are derived within a single approach. Here, in contrast to the approach taken in \cite{Sorensen:2020ygf} (where the contribution of the vector interactions to the energy density is assumed to be a polynomial in baryon density), we derived Eqs.\ \eqref{eq:EOM1}-\eqref{eq:EOM2} using the method of test-particles (see Appendix \ref{app:test-particle_ansatz}), which allows us to chose a completely arbitrary form of the vector single-particle potential $A^{\mu}$; this property will be utilized in Sec.~\ref{sec:parametrization}.

By taking moments of the Boltzmann equation, Eq.\ \eqref{eq:I}, one can derive the conservation laws in the standard way; details of this derivation are given in Appendix \ref{app:energy-momentum_tensor}. In particular, the conserved current density of the $i$-th species is given by
\begin{align}
j_i^{\mu} &= g_i  \int \frac{d^3p}{(2\pi)^3} ~ \frac{\Pi^{\mu}}{ \Pi^0}   ~ \tilde{f}_i ~,
\end{align}
where $g_i$ is the degeneracy, and the net baryon current is defined as
\begin{eqnarray}
j_B^{\mu} \equiv \sum_i B_i j_i^{\mu} ~,
\end{eqnarray}
where $B_i$ is the baryon number of the $i$th species. Meanwhile, the energy-momentum tensor assumes the following form:
\begin{align}
T^{\mu\nu} &= \sum_i g_i \int \frac{d^3\Pi}{(2\pi)^3} ~ \frac{\Pi^{\mu}  \Pi^{\nu}}{\Pi^0} ~ \tilde{f}_i +  A^{\mu}j_B^{\nu}  \nonumber\\
& \hspace{5mm} - ~ g^{\mu\nu}~ \bigg(  n_B U(n_B)  -  \int_0^{n_B} dn' ~ U(n') \bigg) ~,
\end{align}
where we introduce the notation $U(n_B)$ to distinguish the zeroth component vector potential $A^{\mu}$ calculated in the rest frame of the baryon current, $U(n_B) \equiv A^{0}(n_B)\big|_{\substack{\txt{rest} \\ \txt{frame}}} =  \alpha (n_B) n_B$. 
Note, in particular, that in equilibrium $T^{\mu\nu} = \txt{diag} (\mathcal{E},P,P,P)$, giving us access to the thermodynamic properties of the system.

The basis of our approach is the fact that by parametrizing the function $\alpha(n_B)$, introduced in Eq.\ \eqref{eq:vector_field}, one can reproduce given properties of nuclear matter in equilibrium and thus control the EOS of baryons, while at the same time the mean-field interactions which lead to a given EOS also enter the equations of motion, Eqs.\ \eqref{eq:EOM1}-\eqref{eq:EOM2}, which continue to be well-defined for an out-of-equilibrium evolution and are used in the simulations.

\subsection{Parametrization of the mean-field potential by the speed of sound}
\label{sec:parametrization}

In this section, we will use the formalism described above to introduce a parametrization of the mean-field potential $A^{\mu}$ that reproduces a given behavior of the speed of sound squared $c_s^2$ as a function of the baryon density $n_B$.

In equilibrium, the baryon density of the $i$th baryonic species is given by
\begin{align} \label{eq:baryon_density}
n_{B,i} = g_i \int \frac{d^3\Pi}{(2\pi)^3}  ~ \left[e^{\beta \big(\Pi^0 + \alpha(n_B)n_B - \mu_B\big)} + 1 \right]^{-1} ~,
\end{align}
where $\beta = 1/T$ and $\Pi^0 = \sqrt{ \bm{\Pi}^2 + m_i^2}$. 
At zero temperature and in the absence of scalar interactions (which would lead to effective masses $m_i^* < m_i$ and, consequently, smaller energies required to excite more massive species), the only baryon species present in the range of baryon densities relevant to heavy-ion collisions, $n_B \in [0,5]n_0$, are protons and neutrons (nucleons); therefore, to simplify the notation, in the following derivation we will drop the index $i$ and take $g = 4$, $m = m_N = 938~\txt{MeV}$. Note that assuming only nucleons to be present at $T=0$ does not preclude exciting other baryon states at $T>0$; moreover, while within our formalism the EOS is essentially fixed for nucleons at $T=0$, it still displays nontrivial behavior as a function of $T $, and is applicable to complex systems of many baryonic species that inevitably arise in considerations at finite temperature as well as in heavy-ion collisions.

Taking the $T \to 0$ limit and integrating Eq.\ \eqref{eq:baryon_density} over the Fermi sphere leads to
\begin{align} \label{eq:mu_rho_model}
\mu_B(n_B, T = 0) &= \alpha(n_B)n_B + \left[m_N^2 + \left(\frac{6\pi^2  n_B}{g} \right)^{2/3}\right]^{1/2} \,.
\end{align}
At the same time, at $T = 0$ the expression for the speed of sound squared is
\begin{align}
c_s^2(n_B, T = 0) &= \frac{n_B}{\mu_B \left(\parr{n_B}{\mu_B}\right)} ~.
\end{align}
Solving the above differential equation for $\mu_B$ yields
\begin{align}\label{eq:mu_rho_cs2}
\mu_B(n_B, T=0) &= \mu_B\big(n_B^{(0)}\big) \exp \Bigg[\int_{n_B^{(0)}}^{n_B} dn' ~   \frac{c_s^2(n')}{n'} \Bigg] ~,
\end{align}
where $n_B^{(0)}$ is some density at which we know the corresponding value of the chemical potential $\mu_B\big(n_B^{(0)}\big)$. Equating the left-hand sides of Eqs.\ \eqref{eq:mu_rho_model} and \eqref{eq:mu_rho_cs2}, we obtain the single-particle rest frame potential $U(n_B) \equiv A^{0}(n_B)\big|_{\substack{\txt{rest} \\ \txt{frame}}} = \alpha (n_B) n_B $,
\begin{align} 
\hspace{-5mm} U(n_B) &= \mu_B\big(n_B^{(0)}\big) \exp \Bigg[\int_{n_B^{(0)}}^{n_B} dn' ~   \frac{c_s^2(n')}{n'} \Bigg] \nonumber \\
& \hspace{5mm} - ~  \left[m_N^2 + \left(\frac{6\pi^2  n_B}{g} \right)^{2/3}\right]^{1/2} ~. \label{eq:pot_cs2}
\end{align}
This form of $U(n_B)$ allows one to parametrize the vector potential in an intuitive way, that is \textit{via} the density dependence of the speed of sound at zero temperature. 

For simplicity, in our study we choose the following piecewise functional form of $c_s^2(n_B)$:
\begin{align} \label{eq:cs2_piecewise}
    c_s^2(n_B) = \left.  \begin{cases}
      c_s^2(\text{Skyrme}), & n_B < n_1 = 2n_0 \\
      c_1^2, & n_1 < n_B < n_2 \\
      c_2^2, & n_2 < n_B < n_3 \\
      \dots & \\
      c_m^2, & n_m < n_B
    \end{cases} \right.
\end{align}
Such parametrization can be considered as a zeroth order interpolation of an arbitrary function $c_s^2(n_B)$. In the above, we take into account that, for $n_B < 2n_0$, an arbitrary parametrization would be inadequate as in this region of baryon density the potential is already considerably well constrained. Therefore, in the density range $n_B \in [0,2]n_0$, we adopt a polynomial parametrization of the potential often referred to as the Skyrme potential,
\begin{align}
  U_{\txt{Sk}}(n_B) = C_1 \left(\frac{n_B}{\tilde{n}}\right)^{b_1 - 1} + C_2 \left( \frac{n_B}{\tilde{n}}\right)^{b_2 - 1} \,,
\end{align}
where $\tilde{n} = 0.168$ fm$^{-3}$, $C_1 = -209.2$ MeV, $C_2 = 156.5$ MeV, $b_1 = 2$, and $b_2 = 2.35$. The Skyrme potential with these values of the parameters is the default mean field in \texttt{SMASH}, producing a nuclear matter ground state at $n_0 = 0.166~\txt{fm}^{-3}$ with binding energy per nucleon $E_{\txt{bin}} = -15.65$ MeV and a moderate incompressibility of $K_0 = 236.73$ MeV (the small discrepancy between $n_0$ and $\tilde{n}$ is an inaccuracy of the default \texttt{SMASH} parametrization, which, however, does not affect any of our results). We note here that given $n_0$ and $E_{\txt{bin}}$, which do not vary significantly between different approaches, whether the Skyrme EOS is soft or hard is entirely controlled by the value of $K_0$, with small values of $K_0$ for soft EOSs and large values  of $K_0$ for hard EOSs. Unless stated otherwise, we employ this default parametrization of the potential at $n_B < 2n_0$. Above $2n_0$, the potential is controlled by the speed of sound as can be seen in Eq.\ \eqref{eq:cs2_piecewise}, and by substituting this piecewise functional form of $c_s^2$ into Eq.\ \eqref{eq:pot_cs2}, we obtain
\begin{widetext}
\begin{align}
U(n_B) = \left.  \begin{cases}
     U_{\mathrm{Sk}}(n_B)~, & ~~~n_B < n_1 = 2n_0 \\
     \Big[U_{\mathrm{Sk}}(n_1) + \mu^*(n_1)\Big] \left(\frac{n_B}{n_1}\right)^{c_1^2} - \mu^*(n_B)~, &  ~~~n_1 < n_B < n_2 \\
     \Big[U_{\mathrm{Sk}}(n_1) + \mu^*(n_1)\Big] \left(\frac{n_B}{n_k}\right)^{c_k^2} \prod^{k}_{i=2} \left(\frac{n_{i}}{n_{i-1}}\right)^{c_{i-1}^2} - \mu^*(n_B)~, & ~~~n_k < n_B < n_{k+1}
     \end{cases} \right. 
     \label{eq:potential_final}
\end{align}
\end{widetext}
where we denote
\begin{eqnarray}
\mu^*(n_B) \equiv \left[m_N^2 + \left(\frac{6\pi^2 n_B}{g}  \right)^{2/3}\right]^{1/2} ~.
\end{eqnarray}
From Eq.\ \eqref{eq:potential_final}, it is straightforward to obtain $\alpha(n_B)$ which enters the expression for the vector field, Eq.\ \eqref{eq:vector_field}, with the latter in turn entering the equations of motion, Eqs.\ \eqref{eq:EOM1}-\eqref{eq:EOM2}.

\subsection{Details of the simulations}
\label{sec:details_of_the_simulations}

We implement the vector mean field $A^{\mu}$, parametrized by the speed of sound according to Eq.\ \eqref{eq:potential_final}, by modifying the  existing \texttt{SMASH} implementation of the VDF model \cite{Sorensen:2020ygf} which our model generalizes. The VDF approach, where the EOS is parametrized using a polynomial form of $\alpha(n_B)$, is fully relativistically covariant and, in particular, leads to fully covariant equations of motion, Eqs.\ \eqref{eq:EOM1}-\eqref{eq:EOM2} 
\footnote{We note that, while the mean-field equations are fully covariant, the implementation of the Boltzmann equation with a finite number of test particles breaks the covariance (for example, collisions between test particles occur at a finite distance, the mean field is obtained on a grid with a finite spacing, etc.). However, in the the limit of an infinite number of test particles these effects disappear, and full covariance is restored.}.
The generalization of the VDF model presented in this paper is, likewise, fully relativistically covariant. In particular, the VDF vector mean-field used to reproduce the Skyrme potential at $n_B < 2n_0$, as described above, is fully relativistically covariant as well. We stress this because usually Skyrme potentials used in hadronic transport codes are, in contrast to our case, nonrelativistic. One important advantage of using covariant mean fields is that a Lorentz-contacted, boosted nucleus is still in the ground state. In a nonrelativistic treatment, on the other hand, one has to either not allow for Lorentz contraction of the incoming nuclei, which results in overestimating the penetration time, or the Lorentz-contracted colliding nuclei are not in the ground state, thus possibly underestimating the repulsion needed to describe, e.g., flow data.

The calculation of the baryon density, necessary for solving the equations of motion, is performed on a Cartesian lattice with lattice spacing 0.8 fm in $x$, $y$, and $z$ directions. We have tested two different methods of weighing (also called smearing) a test particle's contribution to density at each node of the lattice: Gaussian smearing and triangular smearing, described in Appendix \ref{sec:density_calculation}. In the case of Gaussian smearing, the weight for a given test particle's contribution to density on the lattice is Lorentz-contracted in the direction of that test particle's motion. In the case of triangular smearing, the lattice itself is contracted in the $z$ direction by a factor $\gamma = \infrac{\sqrt{s_{NN}}}{2m_N}$. In Appendix \ref{sec:NT_effects}, we show that the flow observables are independent of the type of smearing used for the density calculation (provided that certain technical details of the density calculation are done correctly; see the appendix for more details). To obtain the results presented in the following, we have used the Gaussian smearing, which had a slightly lower computational cost due to the smaller required number of lattice nodes.

Throughout the simulations we apply a mixed ensembles method, within which we simultaneously simulate $N_{\txt{ens}}$ events (in this context also called ensembles), each with $N_T$ test particles per particle. The test particles can collide with each other only within one ensemble, and their cross sections are reduced by a factor of $N_T$ so that the scattering rate per test particle is preserved. However, the baryon density, the vector mean field, and the estimated distribution function for Pauli blocking are computed based on test particles in all $N_{\txt{ens}}$ ensembles. We note that for a reliable density calculation, the product $N_T N_{\txt{ens}}$ should be sufficiently large; see Appendix \ref{sec:NT_effects}.

It is important to stress here that we simulate multiple hadronic species and their reactions. In Eq.\ \eqref{eq:I}, the index $i$ goes over all particle species implemented in \texttt{SMASH}, that is over around 120 hadronic species (this number does not account for isospin states, i.e., $\pi^+$, $\pi^-$, and $\pi^0$ are counted as one species; antibaryons and baryons are also counted as one species). Most of these species are short-lived hadronic resonances. Masses, width, decay channels, and branching ratios of the resonances are taken from the Particle Data Group compendium \cite{ParticleDataGroup:2014cgo} whenever available. Unknown or poorly known properties were adjusted to better fit the known cross sections \cite{Steinberg:2018jvv}. The collision integral on the right-hand side of Eq.\ \eqref{eq:I} includes elastic and inelastic $2\leftrightarrow 2$ reactions, $1 \leftrightarrow 2$ resonance formations and decays, as well as $2 \to n$ reactions realized as a string excitation and breakup \cite{Mohs:2019iee}. Details about resonance properties and reaction cross sections can be found in \cite{Weil:2016zrk,Steinberg:2018jvv}. 

We note that we do not utilize recently implemented stochastic collision rates \cite{Staudenmaier:2021lrg,Garcia-Montero:2021haa}, resorting instead to a standard geometric collision criterion within which collisions are performed if $d_{ij} < \sqrt{\sigma_{ij}/\pi}$, where $d_{ij}$ is the transverse distance between two particles in their center-of-mass frame, computed at the time of the closest approach of the two particles, and $\sigma_{ij}$ is the cross section. A consequence of this choice is that collisions occur at a non-zero distance between particles. This causes spurious effects described in detail in \cite{Cheng:2001dz}: there, it was shown that the finite interaction range effectively increases the viscosity and, moreover, that it makes hadronic flow dependent on the technical details of the collision implementation, which differ between various hadronic transport codes. This problem is remedied by using a large number of test particles per particle, $N_T$. As shown in \cite{Cheng:2001dz}, different transport codes can yield significantly different results at $N_{T} = 1$, but already at $N_{T} = 32$ they were all in agreement regardless of implementation details. We performed similar tests in our simulations, focusing on the effect of different values of $N_T$ and $N_{\txt{ens}}$ on flow observables; a detailed description of our results can be found in Appendix \ref{sec:NT_effects}. Based on these tests, we use $N_T = 20$ and $N_{\txt{ens}} = 50$, which we find to be large enough to yield simulation results that are independent of both the specific density calculation scheme and specific collision criterion used.

At collision energies in the range $E_{\txt{lab}} \in [1, 10]~ \txt{GeV/nucleon}$, proton production can be substantially influenced by the light nuclei production mechanism employed in the simulation. One can estimate the relative importance
of the light nuclei by comparing their yields at midrapidity to proton yields. While the role of the light nuclei diminishes as one goes from $E_{\txt{lab}} = 2$ to $8 ~ \txt{GeV/nucleon}$ (for example, at $E_{\txt{lab}} = 2 ~ \txt{GeV/nucleon}$ the proton yield at mid-rapidity is around 52 and the deuteron yield is around 22, while at $E_{\txt{lab}} = 8 ~ \txt{GeV/nucleon}$ the proton yield remains approximately the same and the deuteron yield reduces to around around 14 \cite{E895:1999ziy}), it is nevertheless a very substantial effect in the whole energy range explored in our study. 

The situation is different for the flow observables, for which the effects of the light nuclei on the proton flow are smaller. However, they are still non-negligible, see \cite{Mohs:2020awg} for a dedicated study. In this work, to account for these effects, we employ a deuteron production model introduced in \cite{Oliinychenko:2018ugs}, where deuterons are produced dynamically in a chain of reactions $NN \leftrightarrow d'$,  $Nd' \leftrightarrow N d$, effectively reducing to nucleon catalysis $NNN \leftrightarrow N d$; or in $NN \leftrightarrow d'$, $\pi d' \leftrightarrow \pi d$, effectively reducing to pion catalysis $\pi NN \leftrightarrow \pi d$; or in the reaction $NN \leftrightarrow \pi d$. The implementation and the cross sections are a part of the publicly available \texttt{SMASH} 2.1 \cite{smash_version_2.1}, and are described in detail in \cite{Oliinychenko:2018ugs}. The most important deuteron production reaction at the considered collision energies, due to the high relative abundance of nucleons compared to pions, is nucleon catalysis. While in our approach deuterons are dynamically produced and influence the entire duration of the collisions, no light nuclei other than deuterons are produced in our simulations. Both this fact and the uncertainty of the deuteron production model contribute to the total systematic uncertainties in our results due to light nuclei production  (for other approaches to simulating light nuclei, see, e.g., \cite{LeFevre:2019wuj,Ikeno:2016xpr}).

Above, we have highlighted some of the known unknowns, i.e., features of the simulation that contribute to its systematic uncertainty: the implementation of the density calculation (smearing), the collision scheme, and the light nuclei production model. In addition, in Appendix \ref{sec:dependence_on_model_and_technical_details} we further discuss possible effects on the simulation results due to the nucleus initialization model (Appendix \ref{sec:nucleus_initialization}), Pauli blocking (Appendix \ref{sec:Pauli_Blocking}), the absence of isospin-dependent, Coulomb, and momentum-dependent interactions (Appendix \ref{sec:only_vector_potentials}), and the absence of in-medium cross sections (Appendix \ref{sec:in-medium_cross_sections}). Overall, some of these effects are substantial and require further systematic study. A comparative analysis taking into account all these possible effects is beyond the scope of the current study. Within the effects explored in the present work, we find the effects due to the EOS to be of the leading order.

\section{Sensitivity of the flow measurements to the EOS}
\label{sec:sensitivity}

\begin{figure*}
    \centering
    \includegraphics[width=\textwidth]{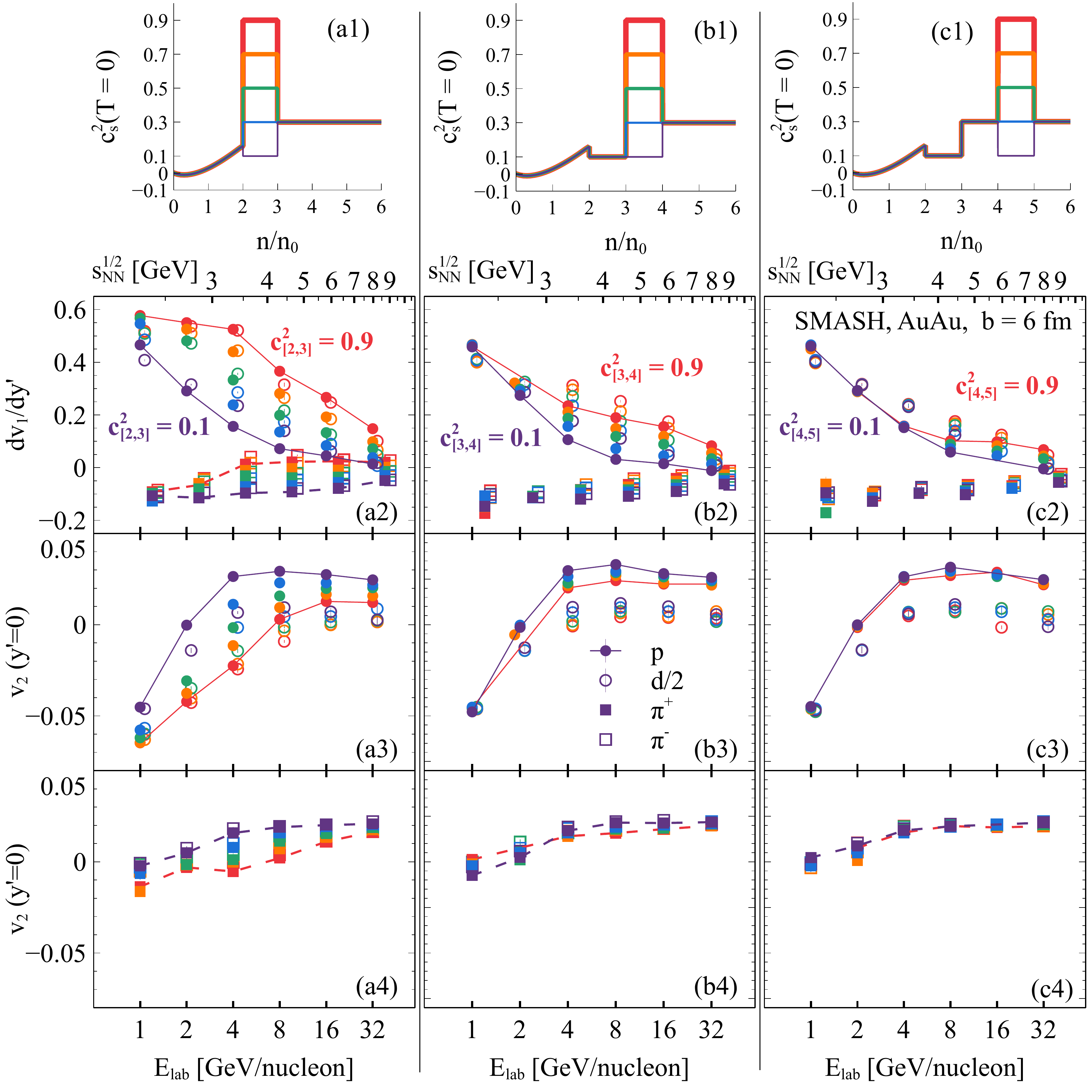}\\
    \caption{Sensitivity of flow observables to the EOS in \texttt{SMASH} simulations of Au+Au collisions at impact parameter $b = 6$ fm. The top panels (a1)-(c1) show  the independent parametrizations of the EOS in three separate density regions; panels (a2)-(c2) show $dv_1/dy'$ for protons (full circles), deuterons (open circles), $\pi^+$ (full squares), and $\pi^-$ (open squares); panels (a3)-(c3) show $v_2(y'=0)$ for protons and deuterons; panels (a4)-(c4) show $v_2(y'=0)$ for pions. Note that the deuteron flow is divided by 2 for both $dv_1/dy'$ and $v_2(y'=0)$, and the points for deuterons and pions are slightly shifted to the right along the $E_{\txt{lab}}$ axis for clarity. Results shown within the columns (a), (b), (c) correspond to different regions in which $c_s^2$ is varied, as can be seen in the top panels; within each column, each EOS (top panels) and the corresponding simulation results (lower panels) are shown using a separate color (purple, blue, green, orange, and red for $c_s^2 = 0.1$, 0.3, 0.5, 0.7, and 0.9, respectively). Note that we show $dv_1/dy'$ instead of $dv_1/dy$, where $y'$ is defined in the center-of-mass frame as $y' = y/y_{\txt{beam}}$, so that $y'_{\txt{beam}} = \pm 1$ in the center-of-mass frame. No $p_T$ cut is imposed.  }
    \label{fig:flow_sensitivity}
\end{figure*}

Our ultimate goal is to constrain the high-density EOS of symmetric nuclear matter using flow measurements in heavy-ion collisions\footnote{We note that while Au nuclei have a finite isospin asymmetry, $\delta \equiv (N_n - N_p)/(N_n + N_p) \approx 0.198$, the asymmetric contribution to the EOS is proportional to $\delta^2 \approx 0.04$ and therefore substantially suppressed. Moreover, while the leading contribution to the symmetry energy is linear in density, the leading contribution to the symmetric EOS is quadratic in density, which makes the symmetry energy contribution less and less important as density increases. Consequently, not only is extracting the EOS of asymmetric nuclear matter considerably challenging at beam energies above $1 ~\txt{GeV/nucleon}$, but also, to a good approximation, at these energies one can neglect the symmetry energy contribution and assume that the extracted EOS is that of symmetric nuclear matter. We make such an assumption in this work.}. In order to do this, in this work we parametrize the EOS using the behavior of $c_s^2(T = 0, n_B)$ as a function of the baryon density (as described in Sec.\ \ref{sec:parametrization}), implement it into the \texttt{SMASH} hadronic transport approach with a vector mean field $A^{\mu} =  \alpha(n_B) j^{\mu}$ (with details of the implementation discussed in Sec.\ \ref{sec:details_of_the_simulations}), simulate heavy-ion collisions, and analyze particle flow from the obtained simulation data. In this section, we set out to investigate two properties of flow observables as used for the extraction of the nuclear matter EOS:
\begin{itemize}
    \item Sensitivity: How much is flow in heavy-ion collisions sensitive to changes of the EOS in a specific density range? 
    \item Specificity: How much is flow in heavy-ion collisions sensitive to other parameters of our simulations (beyond the EOS)?
\end{itemize}

An ideal observable would be sensitive to the EOS at a given density range and insensitive to any other parameters. Additionally, a necessary condition to extract the EOS from experimental data is that variations in simulation results due to the EOS are larger than the experimental errors on the measurements. Our results show (see, e.g., Fig.\ \ref{fig:flow_sensitivity}) that the latter condition is easily satisfied for flow observables, as the change of the flow due to varying the EOS is much larger than the experimental precision. For example, recent STAR measurements of the proton flow at $\sqrt{s_{\mathrm{NN}}} = 3.0$ and 4.5 GeV \cite{STAR:2020dav,STAR:2021yiu} provide $dv_1/dy'$ with an absolute error around 0.01 and $v_2(y'=0)$ with an absolute error around 0.003.

\begin{figure*}
    \centering
    \includegraphics[width=\textwidth]{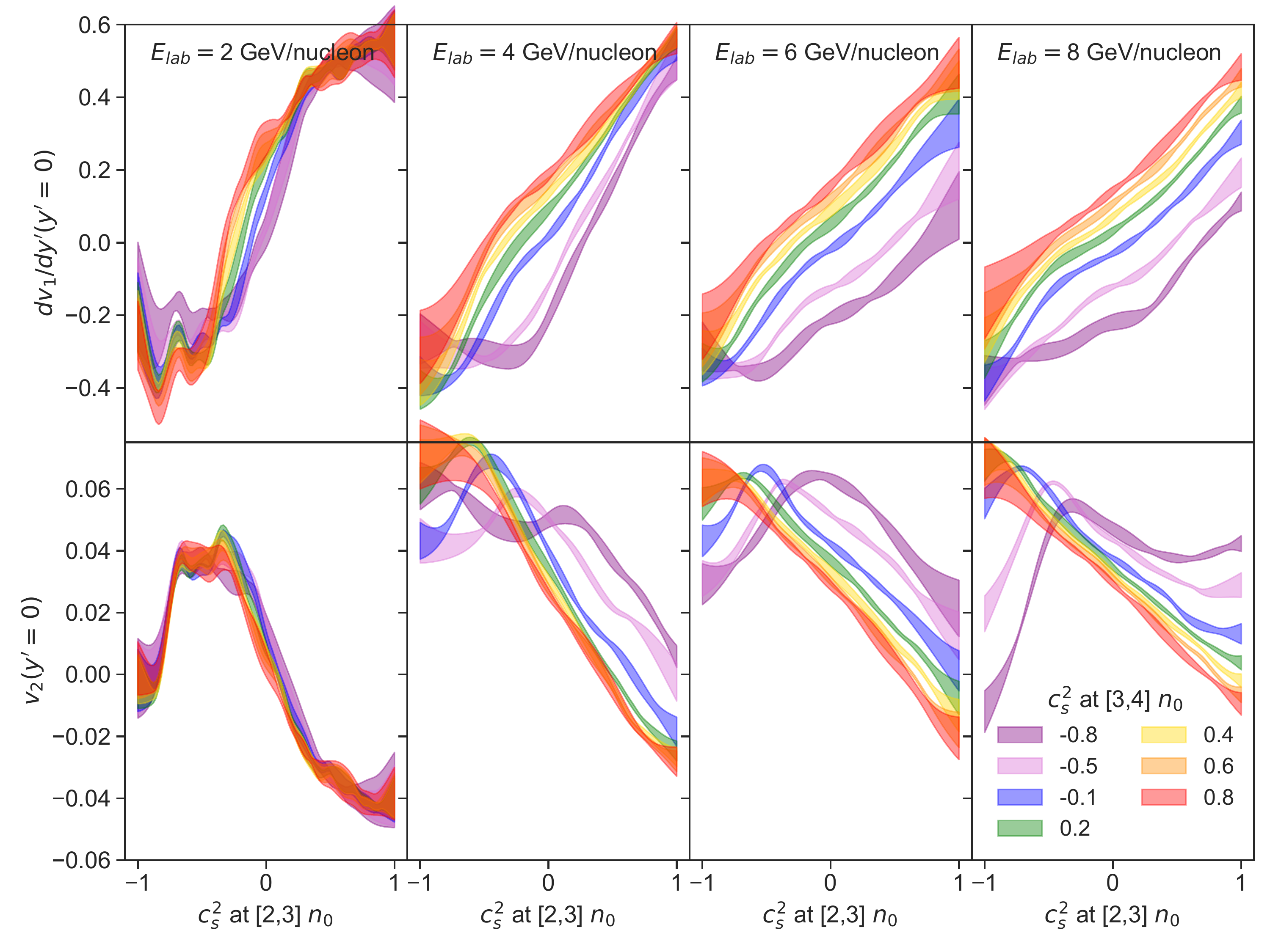}
    \caption{Interpolated dependence of the proton flow in mid-central (with impact parameter $b = 6$ fm) Au+Au collisions on the speed of sound at different densities for a series of collision energies $E_{\txt{lab}}=\{2,~4,~6,~8\} ~\txt{GeV/nucleon}$; see text for more details. }
    \label{fig:flow_vs_cs2}
\end{figure*}

To qualitatively test the sensitivity of flow observables to the EOS in specific density ranges, we perform simulations using a parametrization of the EOS with varying values of $c_s^2(n_B, T=0)$ only in the density range of interest, keeping a fixed dependence on $n_B$ everywhere else, and study how much the flow results depend on the used EOS. In Fig.\ \ref{fig:flow_sensitivity}, we show proton, deuteron, and pion $dv_1/dy'$ and $v_2(y'=0)$ obtained in simulations where $c_s^2(n_B, T=0)$ was varied only in the density range $n_B \in (2,3)n_0$ (first column), only in the density range $n_B \in (3,4)n_0$ (second column), and only in the density range $n_B \in (4,5)n_0$ (third column); for $n_B < 2n_0$, $c_s^2(n_B, T=0)$ took on the density-dependence coming from the underlying default Skyrme EOS (see Sec.\ \ref{sec:parametrization} for more details), while in the remaining, non-varied regions $c_s^2(n_B, T=0)$ took on an arbitrary constant value (see top panels in Fig.\ \ref{fig:flow_sensitivity}).

Figure \ref{fig:flow_sensitivity} clearly shows that in Au+Au mid-central collisions the proton, deuteron, and pion $dv_1/dy'$ and $v_2(y'=0)$ are very sensitive to the EOS at $n_B \in (2, 3)n_0$, with the maximal sensitivity to this density region at the kinetic beam energy around $E_{\txt{lab}} = 4~ \txt{GeV/nucleon}$. It is evident that a stiffer EOS produces a larger $dv_1/dy'$ and a smaller $v_2(y'=0)$ for protons, deuterons, and pions. One can see that the proton and deuteron flow is also sensitive to the density region $n_B \in (3, 4)n_0$, where the maximum sensitivity is reached at the collision energy around $E_{\txt{lab}} = 8~ \txt{GeV/nucleon}$. This sensitivity, however, is smaller than sensitivity to the $n_B \in (2, 3)n_0$ region. The flow of all particles at any collision energy becomes rather insensitive to the EOS for $n_B \in (4, 5)n_0$, which limits the experimental opportunities to constrain the EOS at $n_B > 4n_0$ from heavy-ion collisions (we note, however, that there is a possibility to use the deuteron flow at $E_{\txt{lab}} = 16 ~ \txt{GeV/nucleon}$, where some sensitivity is still present; see Fig.\ \ref{fig:flow_sensitivity}). While one could argue that if the EOS is very soft at lower densities, then the fireball may spend more time at higher densities and, consequently, flow observables might be sensitive to those higher densities, in Fig.\ \ref{fig:flow_sensitivity} we already take a soft EOS at low values of $n_B$ when varying the EOS at $n_B \in (4, 5)n_0$, and the sensitivity at the $n_B \in (4, 5)n_0$ region is still rather small.

Notice that the deuteron flow (for our deuteron production model see Sec. \ref{sec:details_of_the_simulations}), which in Fig.\ \ref{fig:flow_sensitivity} is divided by 2, can be measured with the same experimental precision as the proton flow. Therefore, effectively, deuteron flow is twice more sensitive to the EOS. Pion flow exhibits a moderate dependence on the EOS in the $n_B \in (2, 3)n_0$ region, but at higher densities it becomes rather insensitive to the EOS at any collision energy. Naturally, the most precise constraints can potentially be achieved if one combines experimental data about proton, deuteron, and pion flow.

Let us now assess these sensitivities quantitatively. For this purpose, we simulated mid-central (with impact parameter $b = 6$ fm) Au+Au collisions at collision energies $E_{\txt{lab}}=\{2,~4,~6,~8\} ~\txt{GeV/nucleon}$. The simulations were run for 50 sets of values of $c^2_{[2,3]n_0}$ and $c^2_{[3,4]n_0}$ (design points), sampled randomly in the $(c^2_{[2,3]n_0}, c^2_{[3,4]n_0})$ plane; $c^2_{[4,5]n_0}$ was not varied and instead was set to equal 0.3, based on our observation (see Fig.\ \ref{fig:flow_sensitivity}) that it does not appreciably influence the flow at the considered energies. Next, a Gaussian emulator was used to interpolate between the design points and obtain the flow dependence on the values of $c^2_{[2,3]n_0}$ and $c^2_{[3,4]n_0}$, shown in Fig.\ \ref{fig:flow_vs_cs2}. In this figure, we observe that at the considered energies the dependence of $v_2(y' = 0)$ and especially of $dv_1/dy'$ on $c_s^2$ is very well described by a linear approximation $a_1 + a_2 c^2_{[2,3]n_0} + a_3 c^2_{[3,4]n_0}$, where $a_1,a_2,a_3$ are regression coefficients specific to each energy, as long as $c_s^2$ is positive. Based on this, we can summarize the regression coefficients for protons in Au+Au collisions at $b = 6$ fm as follows:
\begin{align} \label{eq:22}
  \begin{pmatrix}
  \derr{v_1}{y'}|_{2\txt{ GeV}} \\
  \derr{v_1}{y'}|_{4\txt{ GeV}} \\
  \derr{v_1}{y'}|_{8\txt{ GeV}} \\
  v_{2}|_{2\txt{ GeV}} \\
  v_{2}|_{4\txt{ GeV}}  \\
  v_{2}|_{8\txt{ GeV}} 
  \end{pmatrix} \approx &
  \begin{pmatrix}
  0.40 \\
  0.04 \\
   -0.04 \\
    0.00 \\
    0.04 \\
    0.04
  \end{pmatrix} +
  \begin{pmatrix}
  0.35 & 0.00 \\
  0.50 & 0.14 \\
  0.36 & 0.24 \\
  -0.06 & 0.00 \\
  -0.06 & -0.03 \\
  -0.03 &  -0.02
  \end{pmatrix}
  \begin{pmatrix}
  c^2_{[2,3]n_0} \vspace{2mm}\\
  c^2_{[3,4]n_0}
  \end{pmatrix} \,.
\end{align}
We reiterate that the above relations are rather accurate (with the coefficient of determination $R^2 > 0.95$) for positive values of $c_s^2$. Note that the numbers in the second matrix on the right-hand side are effective measures of the sensitivities, i.e., they indicate how well a measurement at a given collision energy constrains the EOS, assuming that there are no other parameters influencing the flow except the EOS. For example, measuring proton $dv_1/dy'$ at $E_{\txt{lab}} = 4 ~ \txt{GeV/nucleon}$ with an error bar of 0.01, one obtains $c^2_{[2,3]n_0}$ with the lowest possible error of around $0.01 / 0.50 = 0.02$, and $c^2_{[3,4n_0]}$ with the lowest possible error of around $0.01 / 0.14 \approx 0.07$. From the same matrix one can also see that $dv_1/dy'$ measurements are more constraining for the EOS, even though $v_2(y'=0)$ is measured with a better precision, because the sensitivity of $v_2(y'=0)$ is lower. Based on both Figs.\ \ref{fig:flow_sensitivity} and \ref{fig:flow_vs_cs2} as well as the sensitivity matrix in Eq.\ \eqref{eq:22}, one can see that the EOS for $n_B \in (2, 3)n_0$ is mainly constrained by measurements at lower energies, around $E_{\txt{lab}} = 4~ \txt{GeV/nucleon}$, while the EOS for $n_B \in (3, 4)n_0$ is mainly constrained by measurements at higher energies, around $E_{\txt{lab}} = 8~ \txt{GeV/nucleon}$.

However, the flow is influenced by parameters other than the EOS parameters. We can evaluate the importance of varying these other parameters (often called nuisance parameters) by finding an equivalent change in the EOS parameters, by which we mean finding how big of an adjustment in, e.g., $c^2_{[2,3]n_0}$ is needed to compensate for the change in flow due to the variation of nuisance parameters. For example, from Fig.\ \ref{fig:xs_sensitivity} one can compute that, at $E_{\txt{lab}} = 4~ \txt{GeV/nucleon}$, scaling all cross sections down by a factor of 0.6 has an effect on $dv_1/dy'$ that can be entirely compensated by increasing $c^2_{[2,3]n_0}$ by around 0.2, as can be seen from Eq.\ \eqref{eq:22}. Such considerations pave the way to a full Bayesian analysis, where both the EOS parameters and the nuisance parameters are varied. However, here we do not perform such an elaborate analysis, given that a computational effort of such a scale should be motivated by a preliminary exploratory study which we provide in this work. Thus we restrict ourselves to varying the EOS parameters and only roughly estimate systematic errors due to the nuisance parameters.

\begin{figure}
    \centering
    \includegraphics[width=0.45\textwidth]{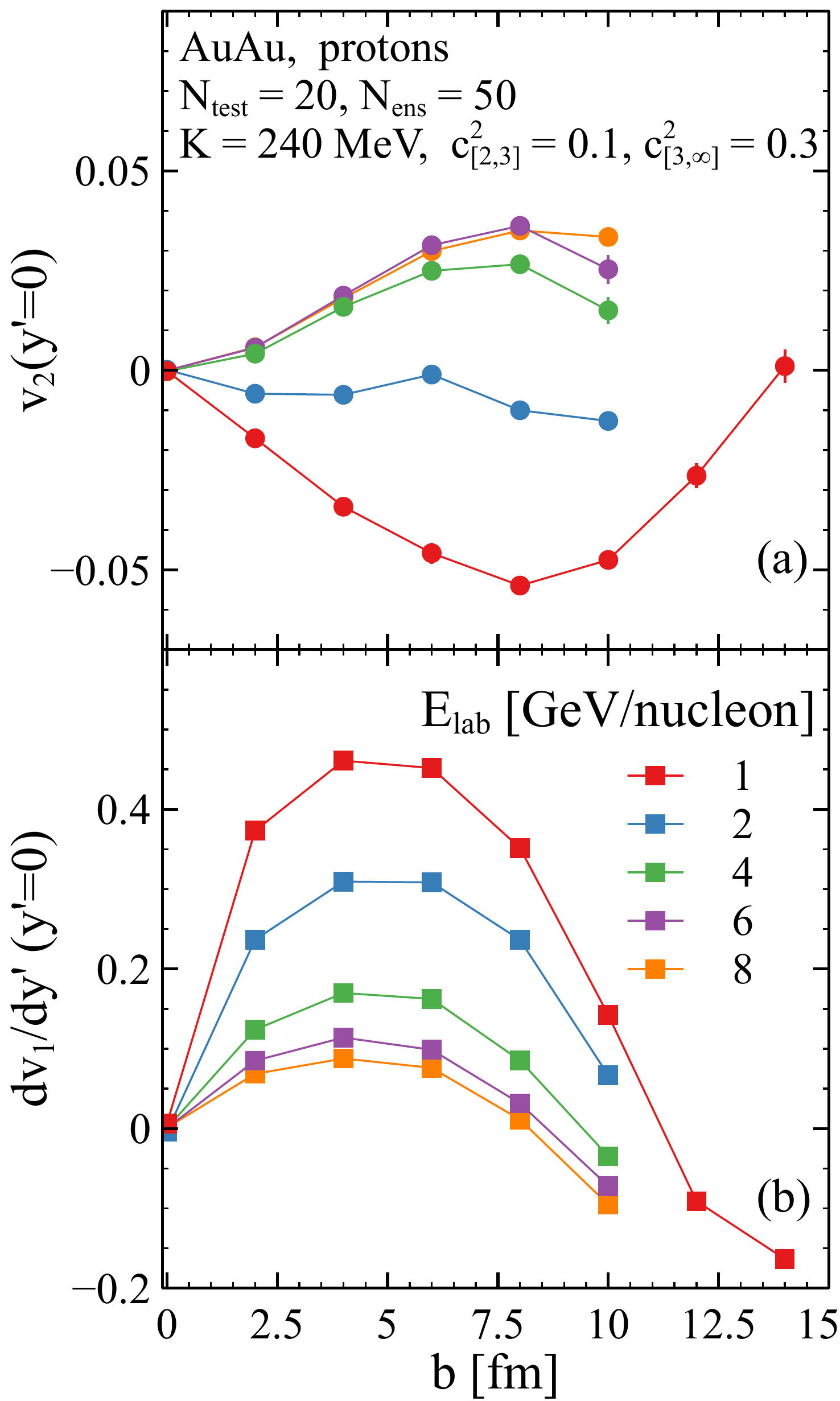}
    \caption{Impact parameter dependence of the proton flow in Au+Au collisions as simulated in \texttt{SMASH}. At $n_B \in (0,2)n_0$, the mean-field is parametrized to reproduce the default Skyrme EOS, while at higher densities the mean-field is parametrized to yield $c^2_s [n_B \in (2,3)n_0] = 0.1$ and $c^2_s [n_B > 3n_0] = 0.3$. Notice that the decrease of $|v_2|(y'=0)$ in peripheral collisions is typical for potentials without explicit momentum-dependent terms, such as the one we employ in this work. With momentum-dependent terms, $|v_2| (y'=0)$ exhibits a monotonic growth against $b$ and attains large values in peripheral collisions; see Fig.\ 4 of \cite{Danielewicz:1999zn}. Which of these scenarios is realized in nature remains to be tested, which will provide strong constraints for the momentum-dependent potential terms.}
    \label{fig:centrality}
\end{figure}

\section{Results}
\label{sec:results}

\subsection{Using both E895 and STAR data}
\label{sec:E895_and_STAR}

Let us start the discussion of our results by reviewing the experimental data that we want to fit to constrain the EOS. We use the flow data from the E895 Collaboration, measured at $E_{\txt{lab}} = \{2,~4,~6,~8\} ~ \txt{GeV/nucleon}$ ($\sqrt{s_{\txt{NN}}} = \{2.7,~ 3.3,~ 3.8,~ 4.3\}~ \txt{GeV}$) \cite{E895:1999ldn,E895:2000maf}, as well as the recent data from the STAR Collaboration at $\sqrt{s_{\txt{NN}}} = \{3, ~ 4.5\} ~ \txt{GeV}$, spanning the collision energy region where the flow is most sensitive to the variation of the EOS above $n_B = 2n_0$ (as already shown in Fig.\ \ref{fig:flow_sensitivity}).  Let us briefly summarize possible systematic differences between measurements from these two experiments. The centrality selection in the E895 experiment was done by considering events with charged particle multiplicities $M$ belonging to a given range of fractions of $M_{\txt{max}}$, where $M_{\txt{max}}$ is the value near the upper limit of the $M$ spectrum at which the height of the distribution has fallen to half of its plateau value. The data presented by the E895 Collaboration are for events with charged particle multiplicity between $0.5 M_{\txt{max}}$ and $0.75 M_{\txt{max}}$, which by comparisons to models was found to correspond to an impact parameter $b = 5$--$7~ \txt{fm}$ \cite{E895:1999ldn,E895:2000maf}. This impact parameter estimation may be imprecise, but we can estimate the error due to a shift in $b$ of $\pm 1~ \txt{fm}$ by using simulation results presented in Fig.\ \ref{fig:centrality}: for example, at $E_{\txt{lab}} = 4 ~ \txt{GeV/nucleon}$, varying $b$ from 5 to 7 fm is equivalent to a decrease in the proton $dv_1/dy'$ of around 0.05, which can be compensated by increasing $c^2_{[2, 3]n_0}$ by around 0.1, as can be seen from Eq.\ \eqref{eq:22}. The STAR centrality selection at $\sqrt{s_{\txt{NN}}} = 3$ GeV is 10--40\%, which is found \cite{HADES:2017def} to correspond to $b = 4.7 $--$ 9.3~ \txt{fm}$. At $\sqrt{s_{\txt{NN}}} = 4.5~ \txt{GeV}$, the centrality selection for calculating $v_1$ is 10--25\%, corresponding to $b = 4.7 $--$7.4~ \txt{fm}$, while for $v_2$ it is 0--30\%, corresponding to $b = 0.0 $--$8.1~ \txt{fm}$.
These differences in centrality selection, along with differences in cuts on the transverse momentum $p_T$, are summarized in Table \ref{tab:exp}.

\begin{figure}
    \centering
    \includegraphics[width=0.45\textwidth]{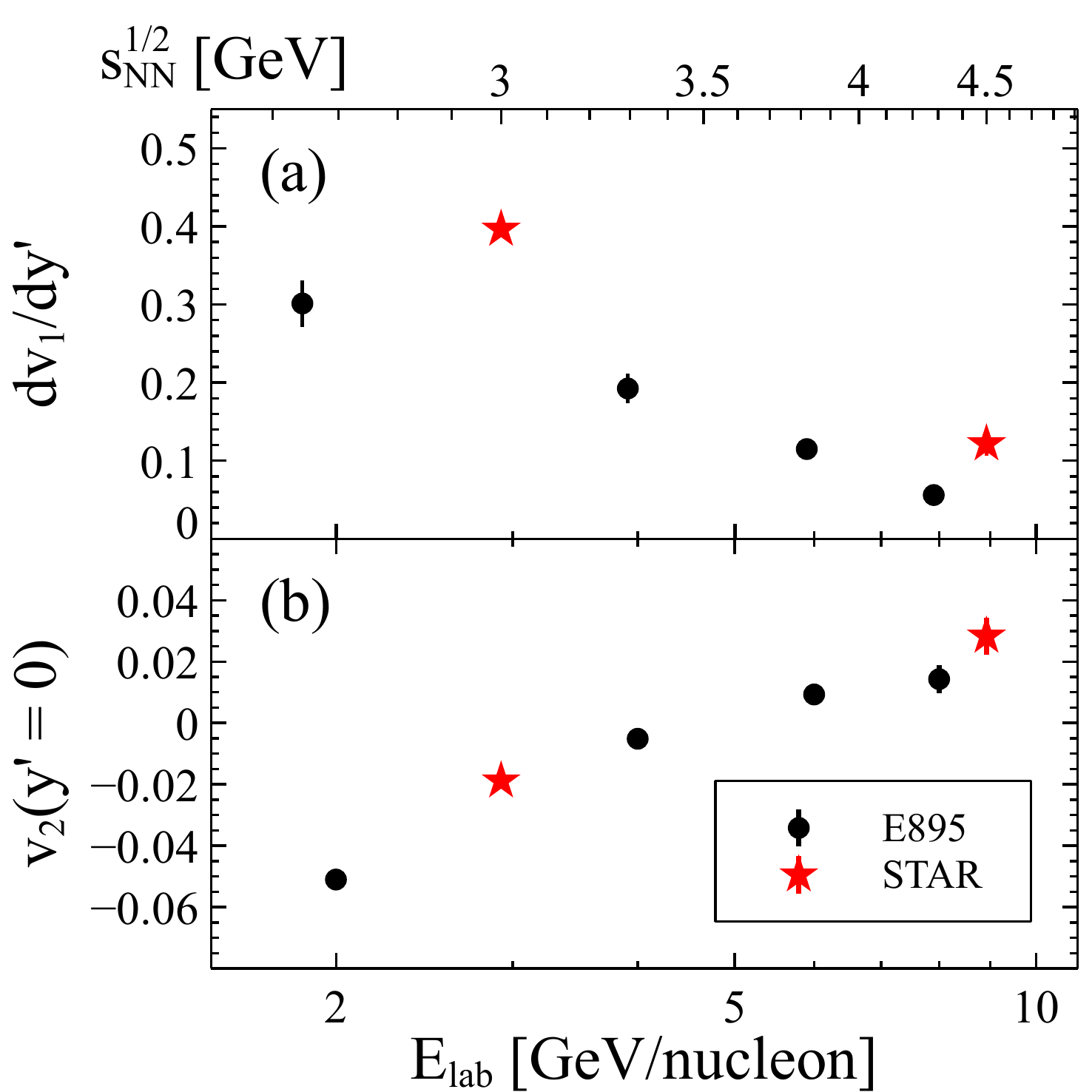}
    \caption{Comparison of E895 \cite{E895:1999ldn,E895:2000maf} and STAR measurements \cite{STAR:2020dav,STAR:2021yiu} for $dv_1/dy'$ (a) and $v_2(y'=0)$ (b) of protons. While the centrality selection and the used $p_T$ cuts are not identical (see Table \ref{tab:exp}), we find that correcting for these factors cannot explain the apparent discrepancy in the $dv_1/dy'$ measurements.}
    \label{fig:flow_expdata}
\end{figure}

\begin{figure}
    \centering
    \includegraphics[width = 0.45\textwidth]{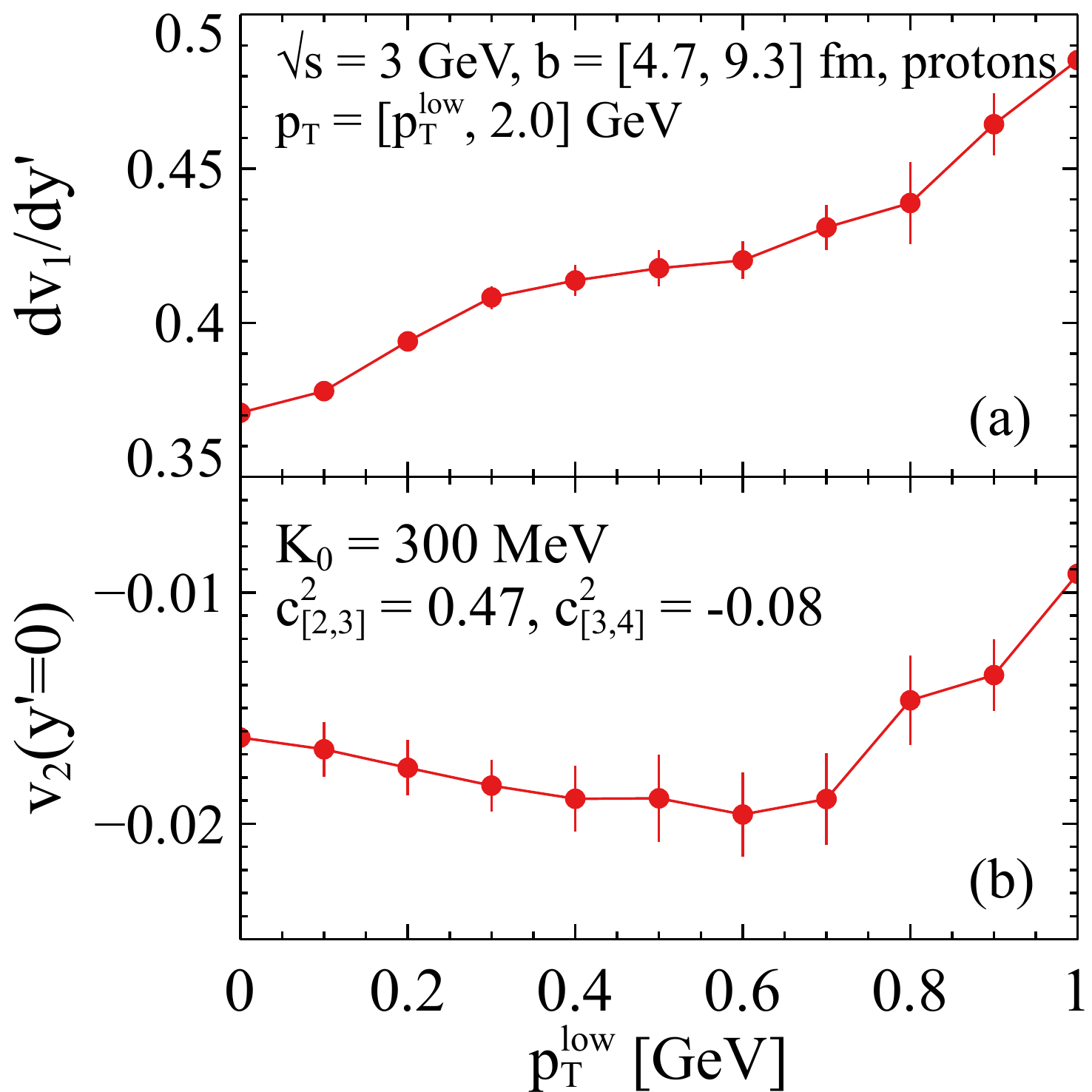}
    \caption{Influence of the lower $p_T$ cut on proton $dv_1/dy'$ (a) and $v_2(y'=0)$ (b) at collision energy $\sqrt{s_{\txt{NN}}} = 3 ~ \txt{GeV}$; we note that in the simulations, we used an EOS parametrized with the maximum \textit{a posteriori} probability parameters $K_0$, $c^2_{[2,3]n_0}$, $c^2_{[3,4]n_0}$, which we will discuss in the following. The shown dependency suggests that the difference between E895 and STAR measurements at $\sqrt{s_{\txt{NN}}} \approx 3 ~ \txt{GeV}$ cannot be ascribed to different $p_T$ cuts applied by the experiments; see Table~\ref{tab:exp}.
    }
    \label{fig:ptcut}
\end{figure}

 Looking at the flow measurements shown in Fig.\ \ref{fig:flow_expdata}, one can see an apparent disagreement between the experiments. It is valid to ask whether they can be explained by a difference in the centrality or $p_T$ selections. We find that they cannot, although to make this conclusion we have to rely on simulations. Our analysis shows that changing the lower $p_T$ bound from 0.4 to 0.1 GeV (as used in the E895 experiment) would lower the STAR $dv_1/dy'$ at $\sqrt{s_{NN}} = 3 ~ \txt{GeV}$, making it closer to the E895 data. However, the magnitude of this change does not exceed 0.05 (see Fig.\ \ref{fig:ptcut}), which is insufficient to explain the discrepancy. Decreasing the upper bound on the centrality selection for STAR at $\sqrt{s_{\txt{NN}}} = 3~ \txt{GeV}$ can only increase $dv_1/dy'$ (see Fig.\ \ref{fig:centrality}), and therefore increase the discrepancy. This makes us conclude that there is a substantial disagreement between the E895 and the STAR flow data, even after taking into the account that they are measured at somewhat different centrality and $p_T$ cuts.

\begin{table}[]
    \centering
    \begin{tabular}{c|c|c|c|c}
    Data & $E_{\txt{lab}}$ [GeV] & $\sqrt{s_{\txt{NN}}}$ [GeV] &  $b$ [fm] & $p_T$-cut [GeV] \\
    \toprule
    E895     & 2   & 2.7   &  5 -- 7  & [0.1, 2.0]   \\
    \hline
    E895     & 4   & 3.4   &  5 -- 7  & [0.1, 2.0]   \\
    \hline
    E895     & 6   & 3.8   &  5 -- 7  & [0.2, 2.0]   \\
    \hline
    E895     & 8   & 4.3   &  5 -- 7  & [0.4, 2.0]   \\
    \hline
    STAR     & 2.9 & 3.0   &  4.7 -- 9.3 & [0.4, 2.0] \\
    \hline
    STAR     & 8.9 & 4.5   & $v_1$: 4.7 -- 7.4 & [0.4, 2.0] \\
     & &   & $v_2$: 0.0 -- 8.10 &
    \end{tabular}
    \caption{Comparison of different experimental conditions for flow measurements \cite{E895:1999ldn,E895:2000maf,STAR:2020dav,STAR:2021yiu}.}
    \label{tab:exp}
\end{table}

\begin{figure*}
    \centering
    \includegraphics[width=\textwidth]{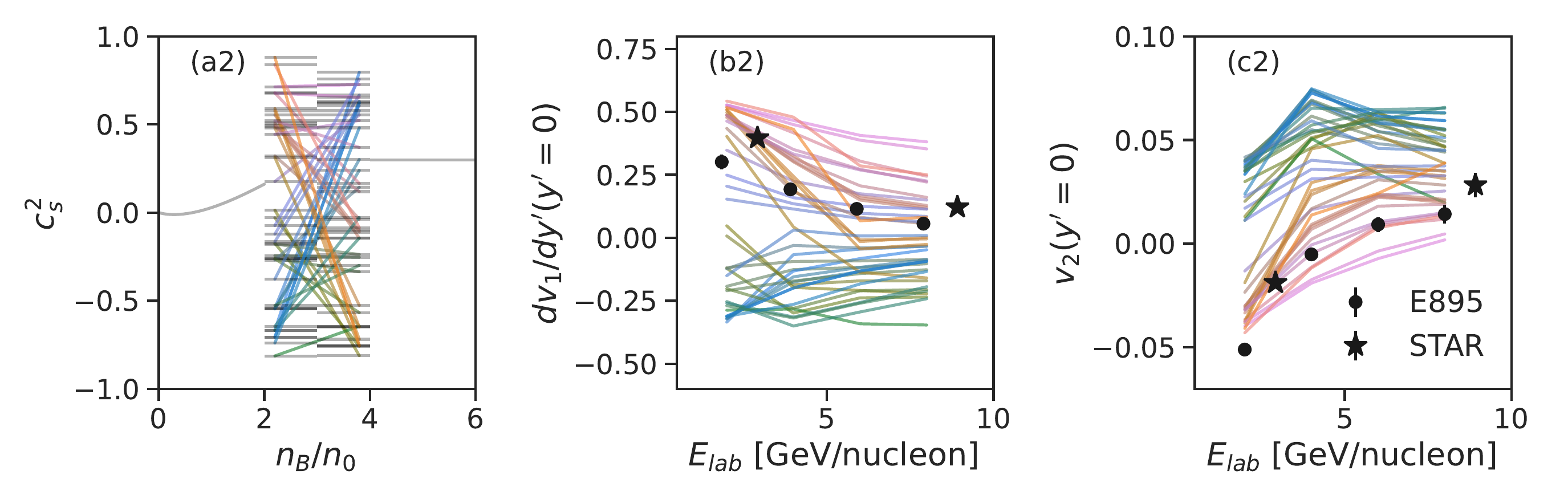}\\
    \includegraphics[width=\textwidth]{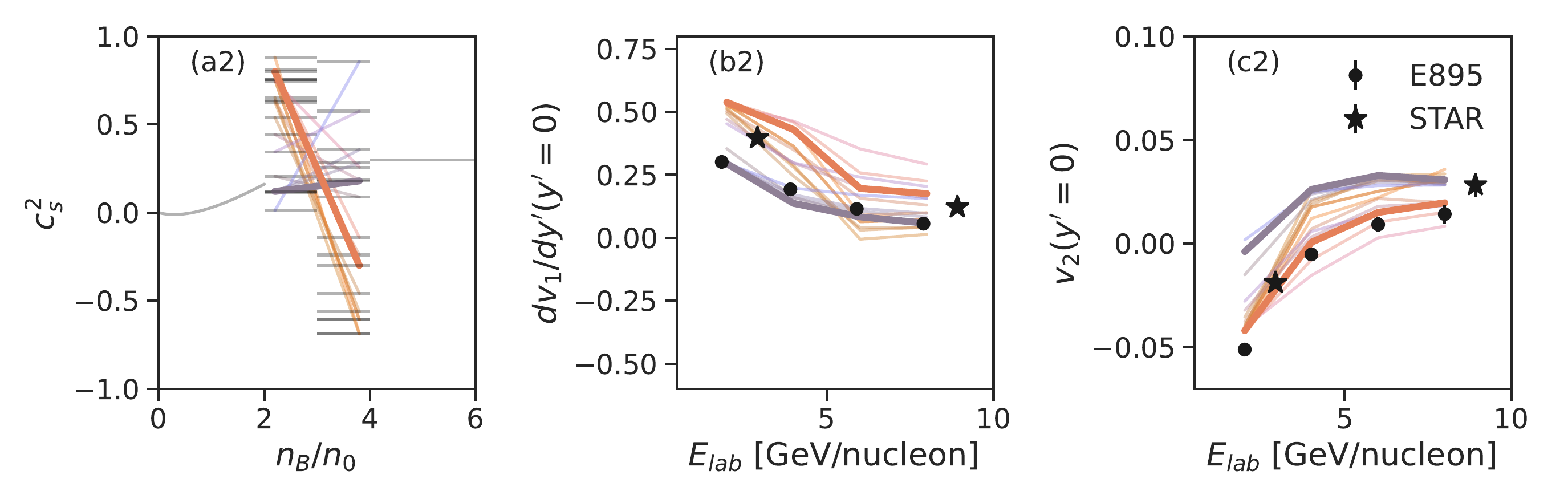}
    \caption{Priors (top) and posteriors (bottom) from analysis employing both E895 and STAR data, where the posteriors were selected by restricting the spread in simulation results for $dv_1/dy'(E_{\txt{lab}})$ and $v_2(E_{\txt{lab}})$ to encompass most of available data. The plot of $c_s^2$ against density in the (a1) panel, showing all EOSs used in the current analysis, has three characteristic regions: for $n_B < 2n_0$, the speed of sound squared follows the behavior given by the default Skyrme EOS; for $n_B > 4n_0$, $c_s^2$ assumes a constant value of 0.3; for $2n_0 < n_B < 3n_0$ and $3n_0 < n_B < 4n_0$, thin horizontal grey lines indicate values of $c_s^2$ in the corresponding region used in the current analysis. The colorful lines joining the horizontal lines for $2n_0 < n_B < 3n_0$ and $3n_0 < n_B < 4n_0$ indicate particular combinations of the values of $c^2_{[2,3]n_0}$ and $c^2_{[3,4]n_0}$ used, with the colors continually changing depending on the values connected. Lines of the same color on all panels correspond to the same combination of $c^2_{[2,3]n_0}$ and $c^2_{[3,4]n_0}$; panels (a1) and (a2) serve as a color legend. By looking at the thick posterior lines, panels (b2) and (c2), one can see that it is not possible, within our model, to fit all experimental data with a single EOS: $v_2$ needs a harder EOS and $dv_1/dy'$ from E895 needs a softer one.}
    \label{fig:bayes_E895_STAR}
\end{figure*}

Given this disagreement in the experimental data, we decide to perform two analyses. First, we try to find a range of the EOS parameters that allows one to roughly encompass both the E895 and the STAR proton flow data. For this purpose we simulate Au+Au collisions at $b = 6$ fm and vary only $c^2_{[2,3]n_0}$ and $c^2_{[3,4]n_0}$, keeping fixed the default Skyrme parametrization at lower densities (see Sec.\ \ref{sec:parametrization} for more details). Both varied parameters are allowed to take any values in the range $c_s^2 \in [-1, 1]$. 

We note here that because a negative speed of sound squared cannot occur in equilibrated matter, regions of the phase diagram where it appears in calculations are often ``corrected'' by performing the Maxwell construction. At the same time, $c_s^2 (n_B, T=0) <0$ simply corresponds to a negative slope of the pressure as a function of density, which occurs when the potential $U(n_B)$ becomes attractive within a certain density range. This in turn indicates that the system is unstable and that, given sufficient time, a phase separation would eventually take place; in fact, it has been shown that such a phase separation indeed occurs in hadronic transport models employing the corresponding mean-field interactions (for a recent study, see \cite{Sorensen:2020ygf} or \cite{Sorensen:2021zxd}). Therefore let us stress, especially in the context of hadronic transport simulations which are well suited for evolving systems out of equilibrium, that there is nothing inherently wrong in an EOS leading to a negative $c_s^2(n_B, T=0)$, as it simply indicates that the particle-particle interactions are on average attractive. Whether a phase separation indeed occurs in that case within heavy-ion collision simulations depends on multiple factors, including the average duration of the collision, the characteristic time for spinodal decomposition, and the average temperature reached in the collision region; nevertheless, regardless of a phase transition occurring or not, the interactions affect the evolution of the system and, consequently, the obtained values of $dv_1/dy'$ and $v_2(y'=0)$.

We start our analysis by assuming prior distributions of the EOS parameters $c^2_{[2,3]n_0}$ and $c^2_{[3,4]n_0}$ to be uniform, and use the degree of agreement between the experimental data and simulations utilizing sampled sets of parameters to infer posterior distributions, that is conditional distributions of parameters given the measurements. In Fig.\ \ref{fig:bayes_E895_STAR}, we show priors (top) and posteriors (bottom) from analysis employing both E895 and STAR data, where the posteriors were selected by restricting the spread in simulation results for $dv_1/dy'(E_{\txt{lab}})$ and $v_2(E_{\txt{lab}})$ to encompass most of the available data. The plot of $c_s^2$ against density, showing all EOSs used in the current analysis (top left panel), has three characteristic regions: (i) for $n_B < 2n_0$, the speed of sound squared follows the behavior given by the default Skyrme EOS for all used EOSs; (ii) for $n_B > 4n_0$, $c_s^2$ assumes a constant value of 0.3 for all used EOSs; (iii) for $n_B \in(2,3)n_0$ and $n_B \in(3,4)n_0$, thin horizontal grey lines indicate values of $c_s^2$ in the corresponding regions that were used in the current analysis, having been sampled uniformly in the respective regions. The colorful straight lines joining the horizontal lines for $n_B \in(2,3)n_0$ and $n_B \in(3,4)n_0$ indicate particular combinations of the values of $c^2_{[2,3]n_0}$ and $c^2_{[3,4]n_0}$ used in a given EOS, with the colors continually changing depending on the values connected; for example, blue lines represent EOSs with considerably negative values of $c^2_{[2,3]n_0}$ and large positive values of $c^2_{[3,4]n_0}$. The same color coding is used in the plots showing flow results for all considered EOSs (top middle and right panels). In the posterior plot for $c_s^2$ (bottom left panel) we emphasize two of the shown EOSs: An orange thick line represents an EOS that is very stiff at $n_B \in(2,3)n_0$ and very soft at $n_B \in(3,4)n_0$, and that can describe $v_2$ from both experiments as well as $dv_1/dy'$ from STAR, but not $dv_1/dy'$ from E895 (see bottom middle and right panels). In contrast, a thick grey line represents an EOS that is moderately soft both at $n_B \in(2,3)n_0$ and at $n_B \in(3,4)n_0$, and that can describe the $dv_1/dy'$ from E895. (We also note here that based on Fig. \ref{fig:bayes_E895_STAR}, one can see that a negative $c^2_{[2, 3]n_0}$ tends to generate $dv_1/dy' < 0$, which is in agreement with a known result from hydrodynamics that a phase transition in the EOS can lead to a negative $dv_1/dy'$ \cite{Stoecker:2004qu}. As we mentioned in the introduction, one can intuitively interpret this as the fireball acting like an attractive spring on the spectators, causing them to be ``deflected inwards''.)

Our broad prior, shown in the upper left panel of Fig.~\ref{fig:bayes_E895_STAR}, covers a large range of proton $dv_1/dy'$ and $v_2(y'=0)$. Even though the substantial disagreement between the E895 and the STAR data makes a precise fit difficult, our posterior distribution restricts possible EOSs to ones with $c^2_{[2,3]n_0} > 0.1$. Nevertheless, the most important conclusion from this analysis is that we cannot simultaneously describe the $dv_1/dy'$ and $v_2(y' = 0)$ measurements obtained by the E895 experiment. The E895 $dv_1/dy'$ data prefer softer potentials, while the E895 $v_2$ data prefer harder ones. This result is in agreement with the conclusion from the seminal work \cite{Danielewicz:2002pu}, where an attempt was made to describe the same data using a different hadronic transport code (known as \texttt{pBUU}, \cite{Danielewicz:1991dh,Danielewicz:1999zn}) and varying the stiffness of the EOS by changing the incompressibility $K_0$.

\begin{figure}
    \centering
    \includegraphics[width=0.49\textwidth]{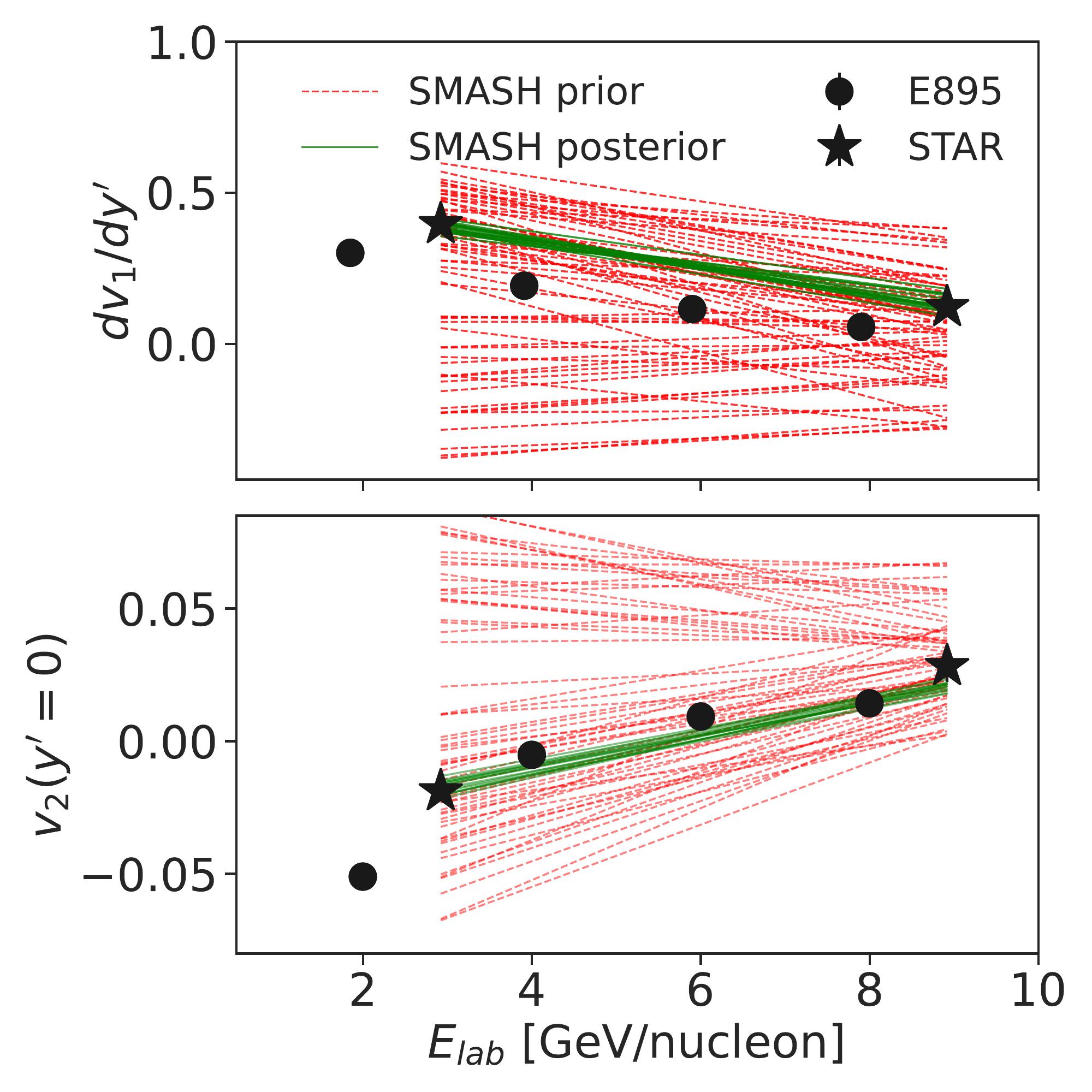}
    \caption{Priors (red dashed lines) and posteriors (blue solid lines) obtained using the STAR proton flow measurements.}
    \label{fig:Bayes_onlySTAR}
\end{figure}

\subsection{Using only STAR data}
\label{sec:only_STAR}

Given the difficulty to simultaneously describe the E895 $dv_1/dy'$ and $v_2(y'=0)$ results, as well as the discrepancy between the E895 and the STAR data, a more rigorous Bayesian analysis with more parameters is unlikely to be helpful if one analyzes both the E895 and the STAR data together. In that case, the final constraint will depend strongly on the systematic errors of the E895 experiment, which are not shown in \cite{E895:1999ldn,E895:2000maf}. Therefore we decided to try discarding the older E895 data from our fit and focus only on the recent STAR proton flow data.  The centrality selection and the $p_T$ cut in our simulations are the same as those for the STAR experiment, listed in Table \ref{tab:exp}. This time, we vary three parameters: the incompressibility $K_0$, which controls the curvature of $\mathcal{E}/n_B$ (and, therefore, $U(n_B)$) at $n_B = n_0$ (for the other two constraints of the Skyrme EOS we take $n_0 = 0.166$ fm$^{-3}$ and $E_{\txt{bin}}$ = -15.65 MeV), $c^2_{[2,3]n_0}$, and $c^2_{[3,4]n_0}$. For $n_B > 4n_0$, $c_s^2$ is fixed at 0.3. To infer the probability distributions of the parameter values, we use the JETSCAPE statistical framework for Bayesian analysis introduced and applied in \cite{JETSCAPE:2020shq}.

\begin{figure}
    \centering
    \includegraphics[width=0.49\textwidth]{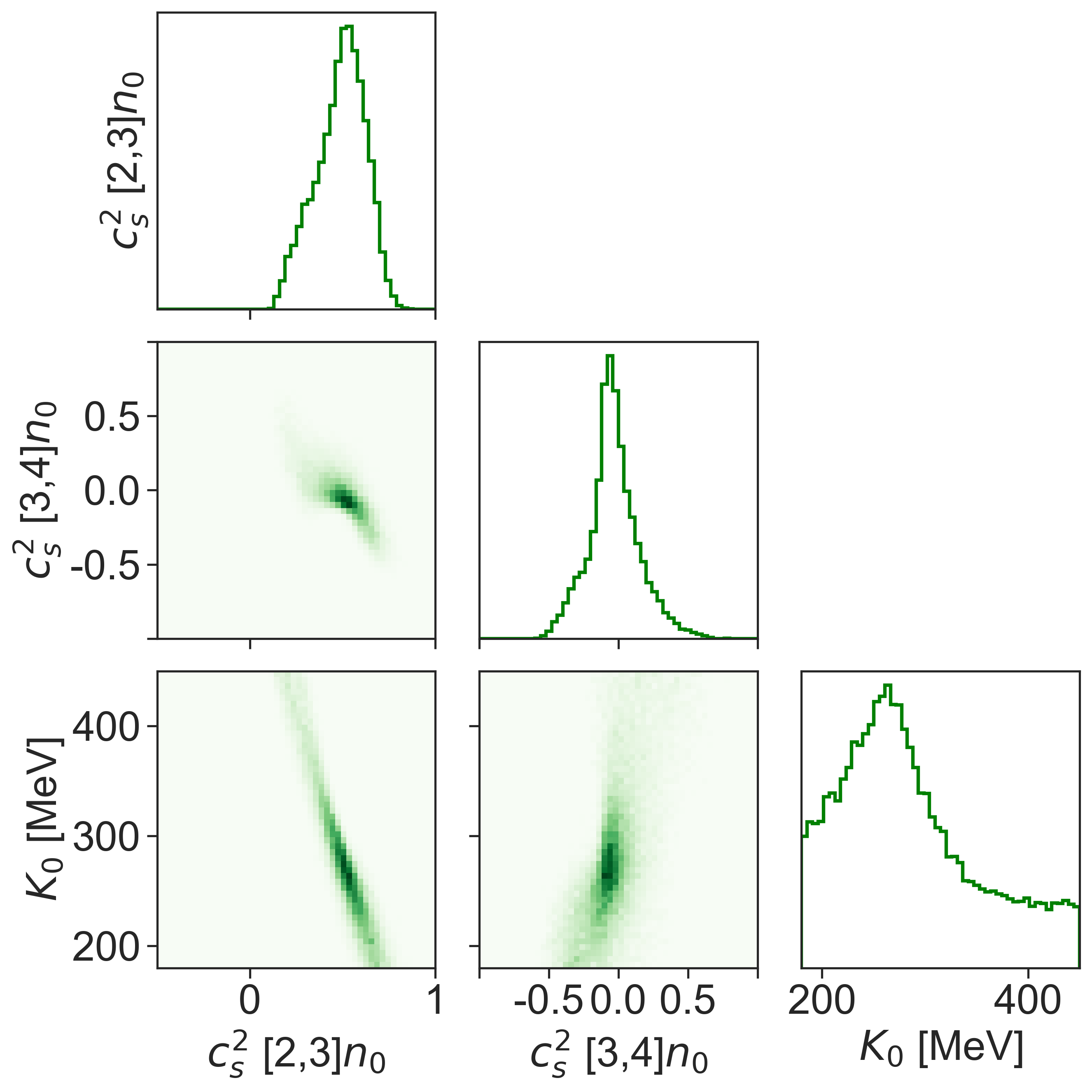}
    \caption{Posterior parameter distribution obtained using the STAR proton flow measurements. The maximum \textit{a posteriori} probability (MAP) parameters are $K_0 = 285 \pm67~\txt{MeV}$, $c^2_{[2,3]n_0} = 0.49 \pm 0.13$, $c^2_{[3,4]n_0} = -0.03 \pm 0.15$.}
    \label{fig:STAR_posterior_parameters}
\end{figure}

In Fig.\ \ref{fig:Bayes_onlySTAR} one can see that, in contrast to the case including the E895 experiment data, our model can fit both the $dv_1/dy'$ and the $v_2(y'=0)$ measurements from the STAR experiment. The posterior distribution of the parameters is shown in Fig.\ \ref{fig:STAR_posterior_parameters}. The incompressibility $K_0$ is not very well constrained by the data, which was expected as the STAR collision energies (and similarly the E895 collision energies) mainly probe densities $n_B > 2n_0$, while a much stronger constraint on $K_0$ can be obtained from lower energy experiments \cite{LeFevre:2015paj} (an extensive discussion of current constraints on $K_0$ can be found in \cite{Dutra:2012mb}; however, see also \cite{Stone:2014wza}). Nevertheless, we can see in Fig.\ \ref{fig:STAR_posterior_parameters} that there is a clear anticorrelation between $K_0$ and $c^2_{[2,3]n_0}$, i.e., that a stiffer EOS at lower densities may be compensated by a somewhat softer EOS at higher densities. Moreover, $c^2_{[2,3]n_0}$ is rather strongly constrained, mainly by the STAR $\sqrt{s_{\txt{NN}}} = 3~ \txt{GeV}$ data, while $c^2_{[3,4]n_0}$ is mainly constrained by the STAR $\sqrt{s_{\txt{NN}}} = 4.5~ \txt{GeV}$ data (see Fig.\ \ref{fig:flow_sensitivity} and the accompanying description in the text).

\begin{figure}[t]
    \centering
    \includegraphics[width=0.49\textwidth]{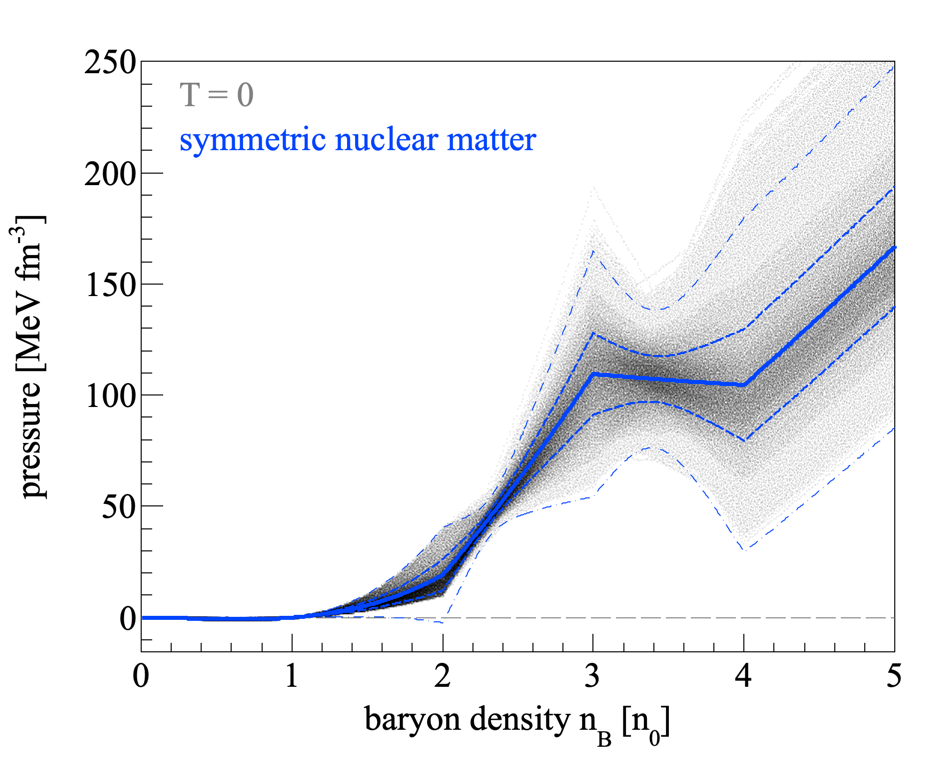}
    \caption{Scatter plot of the pressure as a function of baryon density, obtained with 50000 EOSs with $K_0$, $c^2_{[2,3]n_0}$, and $c^2_{[3,4]n_0}$ sampled from the posterior distribution. The solid line corresponds to the mean of the pressures, while the thick and thin dashed line corresponds to the $\pm 1\sigma$ and $\pm3\sigma$ contour around the mean, respectively.}
    \label{fig:50000_EOSs}
\end{figure}

The maximum \textit{a posteriori} probability (MAP) parameters are $K_0 = 285 \pm67~\txt{MeV}$, $c^2_{[2,3]n_0} = 0.49 \pm 0.13$, $c^2_{[3,4]n_0} = -0.03 \pm 0.15$. Given this result, we conclude that the recent STAR proton flow measurements indicate a very hard potential at $n_B \in (2,3)n_0$, followed by a substantial softening at $n_B \in (3,4)n_0$. Indeed, in Fig.~\ref{fig:50000_EOSs}, showing a scatter plot of the pressure as a function of baryon density for 50000 EOSs obtained by sampling $K_0$, $c^2_{[2,3]n_0}$, and $c^2_{[3,4]n_0}$ from the posterior distribution and putting $c^2_{[4,+\infty]n_0} = 0.3$ (we stress here that it was not possible to constrain the values of $c_s^2$ for $n_B > 4n_0$ within our study, see Sec.\ \ref{sec:sensitivity} for more details), one can see that the pressure prefers a substantial softening for $n_B \in (3,4)n_0$.

A negative $c^2_{[3,4]n_0}$ (or, equivalently, a negative slope of pressure for $n_B \in (3,4)n_0$) is a sign that the EOS exhibits a first-order phase transition in that region. Using the thermodynamics of the model (see Appendix \ref{app:thermo}), one can identify the critical temperature for the EOS with the MAP parameters at $T_c \approx 70\ \txt{MeV}$. At the same time, an EOS with $c^2_{[3,4]n_0} = -0.18$, which is a value smaller by 1$\sigma$ than the central value, corresponds to a critical temperature of $T_c \approx 260\ \txt{MeV}$, while an EOS with the value of $c^2_{[3,4]n_0}$ larger by 1$\sigma$ than the central value, $c^2_{[3,4]n_0} = 0.12$, does not lead to a first-order phase transition at all, although it still displays a significant softening. This is, of course, a very large spread of possible thermodynamic properties of dense nuclear matter in this region. Drawing firm conclusions on the EOS using our study is further complicated by the lack of momentum-dependent interactions in our model. In particular, since at higher energies momentum-dependent interactions are repulsive, describing the experimental results using a model without momentum-dependence leads to a spuriously stiff EOS. If this effect is substantial for $n_B \in (2,3)n_0$, then we can see from Fig.\ \ref{fig:STAR_posterior_parameters} that a very stiff EOS at $n_B \in (2,3)n_0$ can induce a very soft EOS at $n_B \in (3,4)n_0$. Since it is unclear to what extent the stiffness of the EOS for densities $n_B \in (2,3)n_0$ is caused by the the lack of momentum-dependent interactions, we remain cautious about making strong conclusions on the EOS.

\begin{figure}[t]
    \centering
    \includegraphics[width=0.49\textwidth]{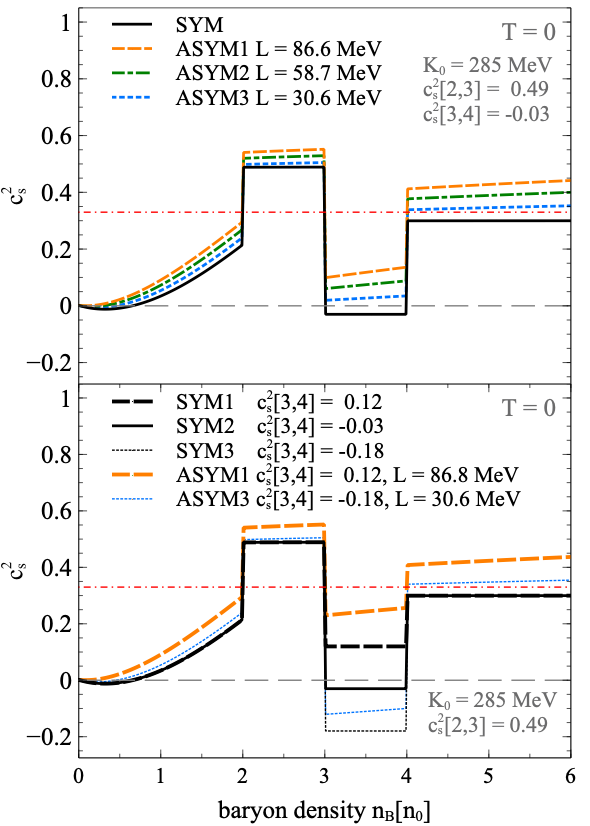}
    \caption{Top panel: The speed of sound $c_s^2$ calculated for the central values of the MAP parameters for both exactly symmetric nuclear matter (solid black line) and pure neutron matter, the latter obtained using three values of the slope parameter of the symmetry energy $L$ (yellow long-dashed, green dash-dotted, and blue short-dashed lines); a smaller (bigger) $L$ results in a smaller (bigger) $c_s^2$ in pure neutron matter. Note that, for all values of $L$ used, the transformation to pure neutron matter results in a disappearance of the first-order phase transition. Bottom panel: The speed of sound $c_s^2$ for exactly symmetric nuclear matter at three values of $c^2_{[3,4]n_0} = \{ 0.12,  -0.03,  -0.18\}$ (thick black long-dashed, medium solid, and thin short-dashed line, respectively), corresponding to the central MAP value and values at $\pm 1 \sigma$, and $c_s^2$ in pure neutron matter obtained using a high value of $L$ for SYM1 (ASYM1, thick yellow long-dashed line) and a low value of $L$ for SYM3 (ASYM3, thin short-dashed blue line). The spread between the ASYM1 and ASYM3 curves illustrates the uncertainty in $c_s^2$ in pure neutron matter at $n_B \in [3,4]n_0$ given both the uncertainty in our results and the uncertainty in the value of $L$.}
    \label{fig:cs2_symmetry_energy_expansion}
\end{figure}

Nevertheless, taking our results at their face value, we see that they imply a substantial softening of the EOS at large baryon densities. One may ask whether this result is consistent with the current knowledge of the EOS based on neutron star data~\cite{Bedaque:2014sqa,Tews:2018kmu,Fujimoto:2019hxv, Marczenko:2022jhl}, which indicates a very stiff EOS at moderate densities, including a local maximum in the speed of sound at which $c_s^2 > 1/3$, and significantly disfavors large softening of the EOS that inevitably arises when a first-order phase transition is present. Importantly, the differences between symmetric (heavy-ion collisions) and asymmetric nuclear matter (neutron stars) should not be overlooked. To test the influence of the symmetry energy on the behavior of $c_s^2$ as a function of density, we take a simple expansion of the symmetry energy around $n = n_0$ (see Appendix \ref{sec:symmetry_energy_expansion} for more details) and use it to transform the speed of sound squared extracted from our analysis, calculated for exactly symmetric nuclear matter, to the speed of sound squared in pure neutron matter. Given the uncertainty in estimates for the value of the symmetry energy slope parameter, we test three values of $L$ corresponding to the central and central $\pm 1\sigma$ values, which, taking $L = 58.7 \pm 28.1 ~ \txt{MeV}$ \cite{Oertel:2016bki} (see also \cite{Li:2013ola}), means that we use $L = \{30.6, ~58.7, ~86.6 \}~\txt{MeV}$. In the top panel of Fig.\ \ref{fig:cs2_symmetry_energy_expansion}, the $c_s^2$ in symmetric nuclear matter, calculated using the central values of the MAP parameters (that is for $K_0 = 285~ \txt{MeV}$, $c^2_{[2,3]n_0} = 0.49$, $c^2_{[3,4]n_0} = -0.03$, and setting $c^2_{[4,\infty]n_0} = 0.3$), is shown with the black solid line. With yellow long-dashed, green dash-dotted, and short-dashed blue lines we show the corresponding results for pure neutron matter, using the three values of the slope parameter of the symmetry energy $L$. It is evident that a smaller (bigger) $L$ results in a smaller (bigger) $c_s^2$ in pure neutron matter. Note that for all values of $L$ used, the transformation to pure neutron matter results in a disappearance of the first-order phase transition. In the bottom panel, we show $c_s^2$ for exactly symmetric nuclear matter at three values of $c^2_{[3,4]n_0} = \{ 0.12,  -0.03,  -0.18\}$ (thick black long-dashed, medium solid, and thin short-dashed line, respectively), corresponding to the central MAP value and boundary values within 1$\sigma$. For the pure neutron matter, we show two curves of $c_s^2$, obtained using a high value of $L$ for SYM1 (ASYM1, thick long-dashed yellow line) and a low value of $L$ for SYM3 (ASYM3, thin short-dashed blue line). The spread between the ASYM1 and ASYM3 curves illustrates the uncertainty in the speed of sound in pure neutron matter at $n_B \in [3,4]n_0$ given both the uncertainty in our results and the uncertainty in the value of the slope parameter $L$. We note here that this spread might be even larger given the large values of the symmetry energy slope parameter reported by the PREXII experiment, $L = 106 \pm 37~ \txt{MeV}$ \cite{PREX:2021umo}. We also point out that recently, an extensive study was performed in which the influence of the symmetry energy expansion parameters on the conversion between neutron matter EOS and symmetric matter EOS was studied in detail \cite{Yao}.

\begin{figure}[t]
    \centering
    \includegraphics[width=0.49\textwidth]{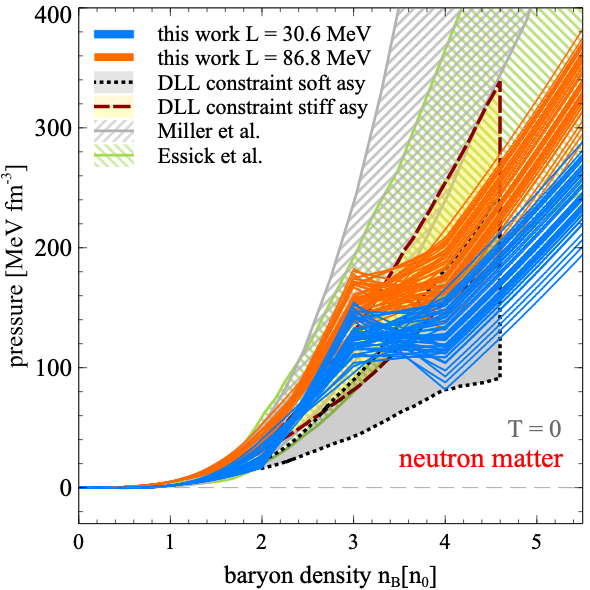}
    \caption{Pressure in pure neutron matter as a function of baryon density for 50 EOSs sampled from the posterior distribution, calculated using two limiting values of L: $L = 30.6 ~ \txt{MeV}$ (blue solid lines) and $L = 86.8 ~ \txt{MeV}$ (orange solid lines). Also shown are the previous constraints based on heavy-ion data from \cite{Danielewicz:2002pu}, likewise calculated for a soft (grey shaded region) and a stiff (yellow shaded region) symmetry energy, as well as a constraint based on the recent NICER analysis of the J0740 pulsar measurement \cite{Miller:2021qha} (area with grey backward stripes) and a constraint based on a global analysis of available neutron star data \cite{Essick:2020flb} (area with green forward stripes).}
    \label{fig:P_vs_n_constraint}
\end{figure}

Based on the above discussion, we see that in general a lack of a first-order phase transition in pure neutron matter does not exclude a first-order phase transition in exactly symmetric nuclear matter, and in particular we conclude that while there is some tension between our results and the neutron star data, the discrepancy is not significant. Given both the obtained statistical uncertainty of our results and the fact that the study we present is minimal, at this point we do \textit{not} claim that there indeed is a first-order phase transition in exactly symmetric nuclear matter around $n_B \in [3,4]n_0$, but rather, we point out that such a possibility is consistent with the range if possible EOSs indicated by our study.

\begin{figure}[t]
    \centering
    \includegraphics[width=0.49\textwidth]{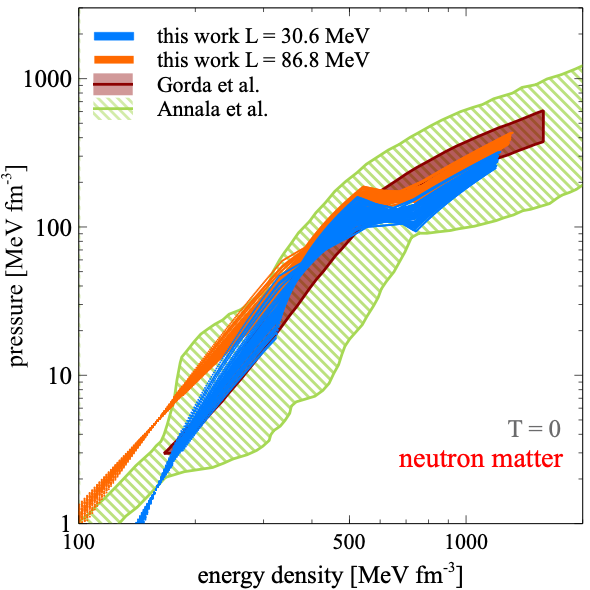}
    \caption{Pressure in pure neutron matter as a function of energy density for 50 EOSs sampled from the posterior distribution, calculated using two limiting values of L: $L = 30.6 ~ \txt{MeV}$ (blue solid lines) and $L = 86.8 ~ \txt{MeV}$ (orange solid lines). Also shown are constraints obtained in two analyses using an interpolation of the EOS at intermediate densities together with perturbative QCD constraints: \cite{Annala:2019puf} (area with green forward stripes) and the more recent \cite{Gorda:2022jvk} (maroon shaded area).}
    \label{fig:P_vs_E_constraint}
\end{figure}

With the symmetry energy expansion, we can explore not only the behavior of the speed of sound, but also of the pressure in pure neutron matter. In Fig.\ \ref{fig:P_vs_n_constraint}, we show pressure in pure neutron matter as a function of baryon density for 50 EOSs sampled from the posterior distribution, calculated using two limiting values of the slope parameter, $L = 30.6 ~\txt{MeV}$ (blue solid lines) and $L = 86.8 ~\txt{MeV}$ (orange solid lines). Also shown are some of the constraints on the pressure from other works: the seminal constraint from \cite{Danielewicz:2002pu} for soft (grey shaded area) and stiff (yellow shaded area) asymmetry energy, a constraint based on an analysis of the NICER measurements of the J0740 pulsar \cite{Miller:2021qha} (area with grey backward stripes), and a constraint from a theoretical analysis utilizing chiral effective field theory coupled with quantum Monte Carlo methods and constrained by observational data on massive pulsars, gravitational waves, and the recent NICER measurement \cite{Essick:2020flb} (area with green forward stripes). It is evident that compared to \cite{Danielewicz:2002pu}, our results favor a much stiffer EOS at lower densities and a somewhat softer EOS at higher densities. The idea that such an EOS would be successful in describing the data was already suggested, although not explored, in \cite{Danielewicz:2002pu}. In particular, we see that the posterior EOSs obtained using a stiff slope parameter are largely consistent with the EOS as inferred by \cite{Essick:2020flb}.

In Fig.\ \ref{fig:P_vs_E_constraint}, we show pressure in pure neutron matter as a function of energy density, where the curves showing the constraint obtained in this work have been obtained in the same fashion as for Fig.\ \ref{fig:P_vs_n_constraint}. Additionally, we show limits obtained from two analyses using an interpolation of the EOS at intermediate densities together with perturbative QCD constraints, \cite{Annala:2019puf} (area with green forward stripes) and the more recent \cite{Gorda:2022jvk} (maroon shaded area). Interestingly, here the EOSs from our softer set are more consistent with the tighter constraint of \cite{Gorda:2022jvk} for energy densities $\mathcal{E} \lesssim 300~ \txt{MeV} \cdot \txt{fm}^{-3}$, or equivalently for densities $n_B \lesssim 2 n_0$, but conversely, the EOSs from the tighter set are more consistent with that constraint at $\mathcal{E} \gtrsim 500~ \txt{MeV} \cdot \txt{fm}^{-3}$, or equivalently for $ n_B \gtrsim 3 n_0$. This suggests an interesting possibility that the symmetry energy could be softer at moderate densities and harder at large densities. 

Finally, let us stress that our results for $n_B \geq 4n_0$ in Figs.\ \ref{fig:50000_EOSs}, \ref{fig:cs2_symmetry_energy_expansion}, \ref{fig:P_vs_n_constraint}, and \ref{fig:P_vs_E_constraint} originate from using $c^2_{[4,\infty]n_0} = 0.3$, a value that we set \textit{ad hoc} due to the lack of a constraint from heavy-ion collisions.

\section{Summary and discussion}
\label{sec:summary}

In this work, we performed a Bayesian analysis of the flow data from heavy-ion collision experiments using hadronic transport simulations. Within our framework, the mean-field potentials can be freely parametrized by the density-dependence of the speed of sound squared at $T=0$ and the incompressibility of nuclear matter, $K_0$. To constrain the EOS, we choose a piecewise parametrization of the speed sound $c_s^2$ in which $c_s^2[n_B<2n_0]$ is that of a Skyrme model (with stiffness controlled by varied values of $K_0$), while $c_s^2[n_B \in (2,3)n_0]$ and $c_s^2[n_B \in (3,4)n_0]$ assume constant values sampled from the interval $[-1,1]$; for $n_B > 4n_0$, we assume $c_s^2[n_B > 4n_0] = 0.3$ (see Sec.~\ref{sec:parametrization} for details). After assessing the sensitivity of the flow measurements to the EOS (Sec.~\ref{sec:sensitivity}), we put a constraint on the dense nuclear matter EOS by performing a Bayesian analysis of simulation results and experimental data (Sec.~\ref{sec:results}). Our study leads us to make the following conclusions: 

\begin{itemize}
    \item Flow observables at $\sqrt{s_{\txt{NN}}} = 2.5$--$5~ \txt{GeV}$ are very sensitive to the dense nuclear matter EOS at $n_B \in (2,4)n_0$. While lower densities can be studied by means of collisions at lower energies, there is almost no possibility to constrain the EOS at baryon densities $n_B > 4n_0$ from AA collisions, at least based on the analysis of flow observables. A similar conclusion was obtained in a concurrent UrQMD study \cite{Steinheimer:2022gqb}.
    \item In particular, we find that the proton flow can yield a very tight constraint on the EOS, and an even better constraint can be obtained from the deuteron flow for which the sensitivity to the EOS is twice as large as that for protons. To a lesser extent, pion flow can also be used to help constrain the EOS. 
    \item Even given a large freedom to vary the EOS differentially in different density regions, we cannot describe the E895 proton flow data: $dv_1/dy'$ prefers a relatively soft EOS while $v_2(y'=0)$ prefers a harder one. However, the more recent STAR data, which additionally seems to be in disagreement with the E895 data, can be described within our model, yielding $K_0 = 285 \pm 67~\txt{MeV}$, $c^2_{[2,3]n_0} = 0.49\pm 0.13$, $c^2_{[3,4]n_0} = -0.03 \pm 0.15$ and thus indicating a very hard EOS at $n_B \in (2,3) n_0$ and a possible phase transition at $n_B \in (3,4) n_0$. 
    \item While the obtained set of EOS parameters results in simulations that reproduce experimental measurements, the extraction of the EOS performed in this work is not definitive given the substantial systematic uncertainties of the model originating from the absence of momentum-dependent potentials. Additional contributions to the uncertainty may come from the choice of the deuteron production model, absence of light nuclei other than deuterons in the model, absence of isospin-dependent and Coulomb interactions, and other sources, including technical details of the simulation.
\end{itemize}

The possibility of a very stiff EOS at relatively low densities above the nuclear saturation density, which our study suggests, is supported by analyses of neutron star EOSs \cite{Bedaque:2014sqa,Tews:2018kmu,Fujimoto:2019hxv, Marczenko:2022jhl}, where the strongest experimental astrophysical constraint comes from the existence of neutron stars with masses above two solar masses \cite{Demorest:2010bx,Antoniadis:2013pzd,Riley:2021pdl,Fonseca:2021wxt}. On the other hand, our finding of a relatively soft EOS at moderate densities, $n_B \in (3,4)n_0$, seems to create some tension with neutron star studies. However, we stress that we only constrain a symmetric nuclear matter EOS. To obtain the neutron star EOS, knowledge of the isospin-dependence of the EOS is necessary; while to some extent one can use the symmetry energy expansion to perform this transformation, at this time the coefficients of the expansion still carry significant uncertainties, and additionally the expansion might only be valid at moderate densities, leaving the behavior of the symmetry energy at high baryon densities largely unknown. Given this uncertainty and the statistical uncertainty from our Bayesian analysis (see Fig. \ref{fig:cs2_symmetry_energy_expansion}) as well as the systematic uncertainty of our model, we see the tension between our results and the neutron star data as not significant. However, it may become meaningful if the mentioned uncertainties decrease.

Our statistical uncertainties can be improved by certain future measurements, assuming that the model will be able to describe them simultaneously. In particular, flow systematics measured at several energies within $\sqrt{s_{\txt{NN}}} = 2.5$--$5~ \txt{GeV}$, that is, the centrality, $y$, and $p_T$ dependences of flow measured for different hadron species (proton, lambda, deuteron, pion, etc.), will decrease statistical errors of the fit, and also allow one to make the EOS parametrization more differential. We also expect that there is a potential for constraining the EOS using measurements of Hanbury-Brown-Twiss (HBT) correlations and dilepton spectra. Describing these measurements will also require improvements of the model, leading to reduced systematic errors.

The systematic errors of our simulations can only be alleviated by improvements of the approach itself. In Sec.~\ref{sec:details_of_the_simulations} and Appendix \ref{sec:dependence_on_model_and_technical_details}, we discuss in detail the known weak points of our model and the possible influence of its technical and theoretical features on the results. In particular, while we found that varying the EOS has a large effect on the simulation results in the studied energy region, other details of the simulation, not constrained within this study, also lead to non-negligible effects, implying substantial systematic errors. Most importantly, momentum-dependent potentials are missing from our approach. While their inclusion is crucial for a differential analysis of flow measurements (e.g., the $p_T$ dependence of $v_2$), their absence in our model introduces a substantial systematic error for integrated flow as well. The next important sources of the systematic error are the deuteron production model as well as not including light nuclei other than deuterons. Furthermore, Coulomb interactions and isospin-dependent potentials were not present in the used model. While their effects seem to be rather small at the studied energies, inclusion of these potentials is required for an accurate assessment of the magnitude of the effect. They would also be required for simulations at lower energies constraining the EOS at densities $n_B< 2n_0$, including meaningfully constraining~$K_0$. Further, some of the incurred systematic errors could be estimated by performing comparisons between different transport simulations, where we note that for the error estimate to be meaningful, it would be important to establish a set of minimum requirements that each such simulation should satisfy, such as sufficiently accurate energy and momentum conservation, reasonable relativistic properties, and reproducing measured particle yields or spectra. Many such comparisons are done within the Transport Model Evaluation Project; see \cite{TMEP:2022xjg} for a review and further references. An overview of multiple potential improvements and research directions in transport modeling was recently provided in \cite{Sorensen:2023zkk}.

In summary, while future improvements on this work are possible both from the theoretical and the experimental side, our current results prove that flow measurements are very sensitive to the underlying EOS at densities $n_B \in (2,4)n_0$ and that, with a better control of the simulations and of the EOS as well as a more expansive Bayesian analysis, they can potentially put a very tight constraint on the EOS of nearly symmetric nuclear matter. Supported by model developments and together with studies of the EOS in neutron stars as well as future tighter constraints on the symmetry energy, heavy-ion data have the power to constrain the EOS of nuclear matter within a sizable region of the QCD phase diagram.

\acknowledgements

Some of results described in this work were presented at the 15th Workshop on Particle Correlations and Femtoscopy at Michigan State University, and we wish to thank the organizers for creating a great atmosphere for scientific discussions. D.O.\ and A.S.\ thank Pawe\l{} Danielewicz, Roy Lacey, Bill Lynch, Betty Tsang, Scott Pratt, Oleh Savchuk, Chun Shen, Sangwook Ryu, Volodymyr Vovchenko, Jan Steinheimer, Nanxi Yao, Jaki Noronha-Hostler, and Veronica Dexheimer for discussions of (selected) results and insightful comments. We also thank the HEPData project, which allowed us to obtain experimental data in a convenient way. D.O.\ thanks Rongrong Ma, Shaowei Lan, Hanna Zbroszczyk, and Declan Keane for comments on experimental data.

D.O., A.S., and L.M.\ acknowledge support by the U.S.\ Department of Energy, Office of Science, Office of Nuclear Physics, under Grant No.\ DE-FG02-00ER41132. V.K.\ was supported by the U.S. Department of Energy, Office of Science, Office of Nuclear Physics, under Contract No.\ DE-AC02-05CH11231.

\appendix

\section{The testparticle ansatz}
\label{app:test-particle_ansatz}

In this appendix we obtain the test-particle equations of motion, Eqs.\ \ref{eq:EOM1}-\ref{eq:EOM2}, by employing the standard test particle ansatz. We base this derivation on a more general case for off-shell particles which can be found in \cite{Leupold:1999ui}.

As already seen in Eq.\ \eqref{eq:III}, the eight-dimensional distribution function $f(x,p)$ can be written as
\begin{eqnarray}
f(x^{\mu},p^{\mu}) = \kappa(\Pi^{\mu}) \tilde{f}(x^{\mu}, \bm{p})~, 
\label{eq:test-particle_ansatz_operational}
\end{eqnarray}
where $\kappa(\Pi^{\mu})$ is a factor ensuring that the momenta in the system are on shell, 
\begin{align}
\kappa(\Pi^{\mu}) = 2 (2\pi) \Theta(\Pi^0) \delta\big(\Pi^{\mu} \Pi_{\mu} - m^2\big) ~,
\end{align}
and the continuous distribution $\tilde{f}$ is approximated by the test-particle ansatz, Eq.\ \eqref{eq:IIIb}, repeated here for convenience with the suppressed index $i$,
\begin{eqnarray}
\hspace{-5mm}\tilde{f}(x, \bm{p}) = \frac{1}{N_T} \sum_{j=1}^{N_T N} \delta^{(3)}\big(\bm{x} - \bm{x}_j(t)\big) \delta^{(3)}\big(\bm{p} - \bm{p}_j(t)\big) ~.
\label{eq:test-particle_ansatz_operational_ftilde}
\end{eqnarray}

Substituting Eq.\ \eqref{eq:test-particle_ansatz_operational} into the Vlasov equation (that is, the Boltzmann equation, Eq.\ \eqref{eq:I}, with the vanishing collision term, $I_{\txt{coll}} = 0$), results in 
\begin{align} 
&\kappa \Big[\Pi_{\mu} \partial^{\mu}_{x}\tilde{f}  + \Pi_{\nu} \big(\partial_{\mu}^x A^{\nu}\big) \partial_p^{\mu}\tilde{f} \Big]  \nonumber \\
& \hspace{10mm} + ~  \tilde{f}\Big[\Pi_{\mu} \partial^{\mu}_{x}\kappa   +  \Pi_{\nu} \big(\partial_{\mu}^x A^{\nu}\big) \partial_p^{\mu}\kappa  \Big] = 0 ~.
\end{align}
We rewrite the expression in the first square bracket by explicitly separating the zeroth components of the four-vectors,
\begin{align}
& \Pi_{0} \partial^{0}_{x}\tilde{f} + \Pi_{\nu} \big(\partial_{0}^x A^{\nu}\big) \partial_p^{0}\tilde{f} \nonumber \\
& \hspace{10mm} + ~\Pi_{i} \partial^{i}_{x}\tilde{f}  + \Pi_{\nu} \big(\partial_{i}^x A^{\nu}\big) \partial_p^{i}\tilde{f}  
~.
\end{align}
From Eq.\ \eqref{eq:test-particle_ansatz_operational_ftilde} it is clear that $\inparr{\tilde{f}}{p^0} = 0$ and $\inderr{\tilde{f}}{t} = 0$, with the latter allowing one to write $\inparr{\tilde{f}}{t}$ as
\begin{align}
\parr{\tilde{f}}{t} &= - \left( \derr{x^k}{t} \parr{}{x^k} + \derr{p^k}{t} \parr{}{p^k}\right) \tilde{f} \nonumber \\
&=  \frac{1}{N_T} \sum_{j=1}^{N_T N}  \left( -\derr{x^k}{t} \partial_k^x - \derr{p^k}{t} \partial_k^p\right)  \nonumber \\
& \hspace{10mm} \times ~\delta^{(3)}(\bm{x} - \bm{x}_j) \delta^{(3)}(\bm{p} - \bm{p}_j)~.
\end{align}
Altogether, the Vlasov equation becomes
\begin{align} 
&\kappa \frac{1}{N_T} \sum_{j=1}^{N_T N} \bigg[ \bigg( \Pi^{i}  - \Pi_{0} \derr{x^i}{t} \bigg) \partial_i^x  \nonumber \\
& \hspace{10mm} + ~ \bigg(-\Pi_{0} \derr{p^i}{t}  + \Pi_{\nu} \big(\partial^{i}_x A^{\nu}\big) \bigg) \partial^p_{i}  \bigg] \nonumber \\
& \hspace{20mm} \times ~\delta^{(3)}(\bm{x} - \bm{x}_j) \delta^{(3)}(\bm{p} - \bm{p}_j) \nonumber \\
& \hspace{10mm} + ~  \tilde{f}\Big[\Pi_{\mu} \partial^{\mu}_{x}\kappa   +  \Pi_{\nu} \big(\partial_{\mu}^x A^{\nu}\big) \partial_p^{\mu}\kappa  \Big] = 0 ~.
\end{align}
The above equation will be always satisfied if, for every test particle $j$, the following sufficient conditions are met:
\begin{align}
& \derr{(x_j)^i}{t} = \frac{(\Pi_j)^{i}}{(\Pi_j)^{0} } ~,
\label{EOM_x}  \\
&  \derr{(p_j)^i}{t}  = \frac{(\Pi_j)^{\nu}}{(\Pi_j)^{0}} \big(\partial^{i}_x A_{\nu}\big) ~, 
\label{EOM_p} \\
& (\Pi_j)^{\mu} (\Pi_j)_{\mu} = m^2 ~,
\label{EOM_on-shell}
\end{align}
where the last condition ensures that $\kappa = \txt{const}$ and, therefore, $\tilde{f}\big[\Pi_{\mu} \partial^{\mu}_{x}\kappa   +  \Pi_{\nu} \big(\partial_{\mu}^x A^{\nu}\big) \partial_p^{\mu}\kappa  \big] =0$. 

To bring Eqs.\ \eqref{EOM_x} and \eqref{EOM_p} into a fully relativistic form, we first note that we can always write (here and in the following we have dropped the index $j$ for clarity)
\begin{eqnarray}
\derr{x^0}{t} = 1 = \frac{\Pi_0}{\Pi_0} ~,
\end{eqnarray}
so that we can immediately write
\begin{eqnarray}
\derr{x^\mu}{t} = \frac{\Pi^{\mu}}{\Pi^{0} } ~,
\label{EOM_x_fully_rel}
\end{eqnarray}
obtaining the same form as given in Eq.\ \eqref{eq:EOM1}.  Similarly, using $p^0 = \Pi^0 + A^0 = \sqrt{\big(\bm{p} - \bm{A} \big)^2 + m^2} + A^0$, we can write
\begin{align}
\derr{p^0}{x_0} &= \frac{\Pi^i}{\Pi_0} \parr{A_i}{x_0} + \parr{A^0}{x_0} 
= \frac{\Pi^{\nu}}{\Pi_0} \parr{A_{\nu}}{x_0} ~,
\end{align}
which can be combined with Eq.\ \eqref{EOM_p} to yield
\begin{equation}
\derr{p^\mu}{t}  = \frac{\Pi_{\nu}}{\Pi^{0}} \big(\partial^{\mu}_x A^{\nu}\big) ~.
\end{equation}
By substituting $p^\mu = \Pi^\mu + A^\mu$ and using
\begin{equation}
\derr{A^{\mu}}{t} = \derr{x_{\nu}}{t} \parr{A^{\mu}}{x_{\nu}}  = \frac{\Pi_{\nu}}{\Pi_0} \big(\partial^{\nu}_x A^{\mu}\big)~,
\end{equation}
we obtain
\begin{equation}
\derr{\Pi^\mu}{t}  = \frac{\Pi_{\nu}}{\Pi^{0}} \Big(\partial^{\mu}_x A^{\nu} -  \partial^{\nu}_x A^{\mu}\Big) ~.
\end{equation}
Finally, defining 
\begin{equation}
F^{\mu\nu} \equiv \partial^{\mu}_x A^{\nu} -  \partial^{\nu}_x A^{\mu}~,
\end{equation}
we obtain
\begin{equation}
\derr{\Pi^\mu}{t}  = \frac{\Pi_{\nu}}{\Pi^{0}} F^{\mu\nu} ~,
\end{equation}
which agrees with Eq.\ \eqref{eq:EOM2}.

It is worth noting that Eqs.\ \eqref{eq:EOM1} and \eqref{eq:EOM2} are analogous to the relativistic equations of motion of a charge in an electromagnetic field, where the crucial difference in our case is that the role of the field is played by the self-consistently calculated baryon four-current instead of an external field. Indeed, it is possible to rewrite Eq.\ \eqref{eq:EOM2} in the form
\begin{align}
\frac{d \bm{\Pi}}{dt} &= \Big[ - \bm{\nabla} A^{0}  - \partial^{0} \bm{A}   \Big] + \frac{\bm{\Pi}}{~\Pi_0} \times \big( \bm{\nabla} \times \bm{A}  \big)   ~,
\label{EOM_VDF_explicit}
\end{align}
further underscoring the analogy (a detailed derivation of the above standard result can be found in Appendix I.4 of \cite{Sorensen:2021zxd}).

\section{Derivation of the four-current and the energy-momentum tensor} 
\label{app:energy-momentum_tensor}

This derivation largely follows \cite{Weber:1992qc}, with two modifications: in our case the vector field $A^{\mu}$ has no explicit momentum dependence, and instead we are dealing with an arbitrary dependence of $A^{\mu}$ on density. 

We start with the BUU equation, Eq.\ \eqref{eq:I}, repeated here in a simplified notation without the index $i$ enumerating the particle species,
\begin{equation} 
\Pi_{\mu} \partial^{\mu}_{x} f + \Pi_{\nu} \big(\partial_{\mu}^x A^{\nu}\big) \partial_p^{\mu} f = I_{\txt{coll}} \,,
\end{equation}
where $f = f(x,p)$. We define $V = \frac{1}{2}\big(\Pi^{\mu}\Pi_{\mu}\big)$; using $\Pi^{\mu} = p^{\mu} - A^{\mu}(x)$, we notice that 
\begin{equation}
\partial_{\mu}^{p} V = \Pi_{\mu}  \hspace{5mm} \txt{and} \hspace{5mm} \partial_{\mu}^x V = - \Pi_{\nu} \big(\partial_{\mu}^x A^{\nu}\big) ~,
\label{eq:derivatives_of_V}
\end{equation}
by means of which we rewrite the BUU equation as
\begin{equation} \label{eq:BUU2}
\big( \partial_{\mu}^p {V} \big) \partial_{x}^{\mu} f - \big(\partial_{\mu}^{x} V \big) \partial_{p}^{\mu}  f = I_{\txt{coll}} \,.
\end{equation}

To obtain the four-current, we integrate Eq.\ \eqref{eq:BUU2} over all possible four-momenta, $g \int d^4p/(2\pi)^4$, where $g$ is the degeneracy factor. The collision integral on the right-hand side of Eq.\ \eqref{eq:BUU2} vanishes due to particle number conservation (an easy demonstration of this fact is shown in Appendix H of \cite{Sorensen:2021zxd}), and we are left with
\begin{align} 
0 &= g \int \frac{ d^4p}{(2\pi)^4} ~\bigg( \big( \partial_{\mu}^p {V} \big) \partial_{x}^{\mu}f- \big(\partial_{\mu}^{x} V \big) \partial_{p}^{\mu} f \bigg) \nonumber \\
&= g \int \frac{ d^4p}{(2\pi)^4}~ \bigg( \big( \partial_{\mu}^p {V} \big) \partial_{x}^{\mu} f +  \Big[\partial_{p}^{\mu} \big(\partial_{\mu}^{x} V \big) \Big]  f\bigg) \nonumber \\
&= g \int \frac{ d^4p}{(2\pi)^4} ~\partial_{x}^{\mu} \Big[ \big( \partial_{\mu}^p {V} \big) ~f \Big] \nonumber \\
&= \partial_{x}^{\mu} \bigg[ g \int \frac{ d^4p}{(2\pi)^4} ~ \big( \partial_{\mu}^p {V} \big) ~ f \bigg] ~,
\end{align}
where in the second and third equalities we used integration by parts and $ab' = (ab)' - a'b $, respectively. Recognizing the above equality as the charge conservation equation $\partial^{\mu} j_{\mu} = 0$, we can identify the four-current as
\begin{align} \label{eq:jmu_8-dimensional}
j_{\mu} & = g \int \frac{ d^4p}{(2\pi)^4} ~ \big( \partial_{\mu}^p {V} \big)~ f ~.
\end{align}
At this point we turn our attention to the fact that the distribution function $f$ is a function of eight independent variables $f = f(x^{\mu}, p^{\mu})$. One integrates over the energy $p^0$ by explicitly including the mass-shell condition, Eq.\ \eqref{eq:III}, in the expression for $f(x,p)$:
\begin{eqnarray}
f(x^{\mu}, p^{\mu})  = 2 (2\pi) \Theta(\Pi^0) \delta\big(\Pi^{\mu} \Pi_{\mu} - m^2\big) ~ \tilde{f}(x^{\mu}, \bm{p}) ~.
\end{eqnarray}
With the identity $\delta\big[ g(x) \big] =\sum_{j} \frac{\delta(x - x_j)}{ |g'(x_j)| }$, where $x_j$ are roots of $g(x)$, the mass-shell condition allows one to write
\begin{eqnarray}
&& g \int \frac{d^4\Pi}{(2\pi)^4} ~ f(x^{\mu}, p^{\mu}) 
=g \int \frac{d^3p}{(2\pi)^3} ~ \frac{1}{ \Pi^0} ~ \tilde{f}(x^{\mu}, \bm{p}) ~,
\label{eq:mass-shell_derived}
\end{eqnarray}
where $\Pi^0 = \sqrt{\bm{\Pi}^2 + m^2}$. Consequently, additionally using Eq.\ \eqref{eq:derivatives_of_V}, we can also write the conserved current in the more well-known form
\begin{align} \label{eq:jmu}
j^{\mu} &= g \int \frac{d^3p}{(2\pi)^3} ~ \frac{\Pi^{\mu}}{ \Pi^0}   ~ \tilde{f} ~.
\end{align}

The energy-momentum tensor is obtained in a similarly standard way, that is \textit{via} multiplying Eq.\ \eqref{eq:BUU2} by $p^{\mu}$ and integrating over $g \int d^4p/(2\pi)^4$. The integral over $I_{\txt{coll}}$ again vanishes, this time due to the conservation of four-momentum in collisions (see Appendix H of \cite{Sorensen:2021zxd} for a simple demonstration), and one obtains
\begin{align}
0 &= g \int \frac{d^4p}{(2\pi)^4} ~  \bigg(p^{\nu} \big( \partial_{\mu}^p {V} \big) \partial_{x}^{\mu} f - p^{\nu}\big(\partial_{\mu}^{x} V \big) \partial_{p}^{\mu}  f \bigg) \nonumber \\
&= g \int \frac{d^4p}{(2\pi)^4} ~  \bigg( \partial_{x}^{\mu} \Big[p^{\nu} \big( \partial_{\mu}^p {V} \big) f \Big] - \Big[\partial_{x}^{\mu} p^{\nu} \big( \partial_{\mu}^p {V} \big) \Big]f \nonumber \\
& \hspace{15mm} +~ g^{\mu\nu} \big(\partial_{\mu}^{x} V \big)  f +  p^{\nu} \Big[\partial_{p}^{\mu}\big(\partial_{\mu}^{x} V \big) \Big]  f \bigg) \nonumber  \\
&= \partial_{x}^{\mu} ~ g \int \frac{d^4p}{(2\pi)^4} ~  \big(\Pi^{\nu} + A^{\nu}\big) \big( \partial_{\mu}^p {V} \big) ~ f \nonumber \\
& \hspace{15mm}  + ~ g^{\nu\mu}~ g \int \frac{d^4p}{(2\pi)^4} ~  \big(\partial_{\mu}^{x} V \big) ~ f ~,
\label{eq:Tmn_start}
\end{align}
where in the first equality we again used $ab' = (ab)' - a'b $ and integration by parts, and in the second equality we substituted $p^{\nu} = \Pi^{\nu} + A^{\nu}$. Because $A^{\nu}$ does not depend on momentum, we can pull it out of the integrals, which together with Eq.\ \eqref{eq:derivatives_of_V} and the  expression for the conserved four-current, Eqs.\ \eqref{eq:jmu_8-dimensional} and \eqref{eq:jmu}, yield
\begin{align}
&  \partial_{x}^{\mu} \Bigg[ g \int \frac{d^4p}{(2\pi)^4} ~ \Pi^{\nu}  \Pi_{\mu} ~ f +  A^{\nu}j_{\mu} \Bigg] \nonumber \\
& \hspace{15mm} - ~ g^{\nu\mu}~ j_{\lambda} \big(\partial_{\mu}^x A^{\lambda}\big)  = 0~.
\label{Tmunu_total_der}
\end{align}
To extract the energy-momentum tensor from the above expression, one needs to write the last term as a total derivative. Recalling that $A^{\lambda} = \alpha(n) j^{\lambda}$ (where we suppress the baryon charge index $B$) and $j_{\mu} j^{\mu} = n^2$, we can write
\begin{align}
j_{\lambda} \big(\partial_{\mu}^x A^{\lambda}\big) 
&= j_{\lambda}  \Big[\partial_{\mu}^x  \alpha(n) \Big] j^{\lambda}  + j_{\lambda} \alpha(n) \Big[ \partial_{\mu}^x  j^{\lambda} \Big] \nonumber \\
&= n^2 \Big[\partial_{\mu}^x  \alpha(n) \Big]  +  \alpha(n) \bigg( \frac{1}{2}  \partial_{\mu}^x \big[ n^2 \big]\bigg) \nonumber \\
&=  \partial_{\mu}^x   \Big[\alpha(n) n^2 \Big]  -  \frac{1}{2} \alpha(n) \Big(   \partial_{\mu}^x \big[ n^2 \big]\Big) ~. 
\label{Tmunu_total_der_second_term}
\end{align}
The second term can be further rewritten as
\begin{align}
\alpha(n) n  \Big[   \partial_{\mu}^x n\Big]  &=  \bigg[\parr{}{n} \int_0^n dn' ~ \alpha(n') n' \bigg]  \bigg[  \parr{n}{x^{\mu}} \bigg] \nonumber \\
&= \parr{}{x^{\mu}} \int_0^{n} dn' ~ n' \alpha (n') ~.
\end{align}
Together with Eq.\ \eqref{Tmunu_total_der_second_term}, the above allows one to write Eq.\ \eqref{Tmunu_total_der} as
\begin{align}
&  \partial^{x}_{\mu} \Bigg[ g \int \frac{d^4p}{(2\pi)^4} ~ \Pi^{\nu}  \Pi^{\mu} ~ f +  A^{\nu}j^{\mu}  \nonumber \\
& \hspace{5mm} - ~ g^{\nu\mu}~  \bigg(  \alpha(n) n^2  -  \int_0^{n} dn' ~ n' \alpha (n')   \bigg) \Bigg] = 0~.
\label{Tmunu_final_der_term}
\end{align}
Because the energy-momentum tensor $T^{\mu\nu}$ satisfies $\partial^x_{\nu} T^{\mu\nu} = 0$, we can immediately identify
\begin{align}
T^{\mu\nu} &=  g \int \frac{d^4p}{(2\pi)^4} ~ \Pi^{\mu}  \Pi^{\nu} ~ f +  A^{\mu}j^{\nu}  \nonumber \\
& \hspace{5mm} - ~ g^{\mu\nu}~ \bigg(  n U(n)  -  \int_0^{n} dn' ~ U(n') \bigg) ~,
\end{align}
where we substituted the expression for the single-particle rest-frame potential as defined in Eq.\ \eqref{eq:potential_final}. By using the mass-shell relation, Eq.\ \eqref{eq:mass-shell_derived}, substituting $d^3p \to d^3 \Pi$ (which is allowed because $A^{\mu}$ does not depend on momentum), and further using the fact that $\Pi^{\mu} = p^{\mu} - A^{\mu}$ and the conserved current $j^{\mu}$ is given by Eq.\ \eqref{eq:jmu}, one finally arrives at
\begin{align}
T^{\mu\nu} &= g \int \frac{d^3\Pi}{(2\pi)^3} ~ \frac{\Pi^{\mu}  \Pi^{\nu}}{\Pi^0} ~ \tilde{f} +  A^{\mu}j^{\nu}  \nonumber\\
& \hspace{5mm} - ~ g^{\mu\nu}~ \bigg(  n U(n)  -  \int_0^{n} dn' ~ U(n') \bigg) ~.
\label{eq:Tmunu_final}
\end{align}
The derived $T^{\mu\nu}$ agrees term by term with expressions in \cite{Sorensen:2020ygf}, where they were derived in the special case of a polynomial form of $\alpha(n)$.

\section{Thermodynamics and thermodynamic consistency}
\label{app:thermo}

One can observe that the energy-stress tensor $T^{\mu\nu}$, derived in Eq.\ \eqref{eq:Tmunu_final}, is symmetric in any frame, and in the Eckart frame, defined as the frame in which $j^{\mu} = (n,0,0,0)$, it becomes diagonal, $T^{\mu\nu} = \txt{diag}(\mathcal{E},P,P,P)$. In that frame, Eq.\ \eqref{eq:jmu} becomes the well-known expression for the rest frame density,
\begin{equation}
n = g \int \frac{d^3p}{(2\pi)^3}   ~ \tilde{f} ~.
\label{eq:TD1}
\end{equation}
Up to this point, the form of the distribution function $\tilde{f}$ was not in any way constrained (and, in particular, it could have been assumed to be given by the test-particle ansatz, Eq.\ \ref{eq:IIIb}, making our formalism perfect for use in hadronic transport). However, because now we are describing a system of fermions in equilibrium, the distribution function $\tilde{f}$ can be assumed to take the well-known form of the Fermi-Dirac distribution, where the single particle energy includes the contribution from the mean-field potential $U(n)$ (see Appendix 1.C in \cite{Sorensen:2020ygf} for a quick derivation, and \cite{BaymPethickLandauFermiLiquidTheory} for a more complete discussion),
\begin{eqnarray}
\tilde{f} = \left( e^{\beta \big(\Pi^0 + U(n) - \mu\big)} + 1 \right)^{-1} ~.
\label{eq:TD2}
\end{eqnarray}

Furthermore, in the Eckart frame the field assumes the simple form $A^{\mu} = \delta^{\mu}_0 \alpha(n) n = \delta^{\mu}_0 U(n)$, and the energy density $\mathcal{E}$ and pressure $P$ are given by
\begin{align}
\hspace{-1mm}\mathcal{E} &= g \int \frac{d^3 \Pi}{(2\pi)^3}  ~ \Pi^0 ~ \tilde{f}   + \int_0^{n} dn' ~  U(n')  ~,
\label{eq:TD3}\\
\hspace{-1mm}P &= g \int \frac{d^3\Pi}{(2\pi)^3} ~ \frac{ \bm{\Pi}^2}{3\Pi^0} ~ \tilde{f} +   n U(n)  -  \int_0^{n} dn' ~ U(n')  ~.
\label{eq:TD4}
\end{align}

Equations \eqref{eq:TD1}-\eqref{eq:TD4} fully describe the thermodynamics of our model, where the parametrization of $U(n)$ can be chosen arbitrarily. The model is thermodynamically consistent, and the system will evolve corresponding to the test particle equations of motion, Eqs.\ \eqref{eq:EOM1}-\eqref{eq:EOM2}.

We note that one can introduce a ``shifted'' chemical potential,
\begin{equation}
\mu^* \equiv \mu - U(n) ~,
\label{eq:effective_mu}
\end{equation}
known as the effective chemical potential. With $\mu^*$ at hand, it is straightforward to see that the density, energy density, and the pressure can be rewritten in the following way:
\begin{align}
n &= n_{\txt{id}}(T,\mu^*)~, 
\label{eq:TD1_vol2} \\
\mathcal{E} &= \mathcal{E}_{\txt{id}}(T, \mu^*)  + \int_0^{n} dn' ~  U(n')  ~,
\label{eq:TD3_vol2}\\
P &= P_{\txt{id}}(T, \mu^*)  +   n U(n)  -  \int_0^{n} dn' ~ U(n')  ~,
\label{eq:TD4_vol2}
\end{align}
where $n_{\txt{id}}$, $\mathcal{E}_{\txt{id}}$, and $P_{\txt{id}}$ are the density, energy density, and pressure of an ideal Fermi gas with chemical potential $\mu^*$, and the total energy density and pressure are obtained by adding the field contributions. In particular, we can also see that the entropy density is given by
\begin{align}
s(T,\mu) &\equiv \frac{1}{T} \Big( \mathcal{E} + P - \mu n \Big) \nonumber \\
&=  \frac{1}{T} \Big( \mathcal{E}_{\txt{id}}(T, \mu^*)  + P_{\txt{id}}(T, \mu^*)  - \mu^* n   \Big)~,
\label{eq:entropy_density}
\end{align}
which is the entropy density of an ideal Fermi gas at the effective chemical potential $\mu^*$.

As a first test of the thermodynamic consistency of Eqs.\ \eqref{eq:TD1_vol2}-\eqref{eq:TD4_vol2}, we check whether the thermodynamic relation $\inbparr{P}{\mu}_T = n$ is fulfilled. Integration by parts allows one to rewrite the ideal-Fermi-gas--like contribution to the total pressure,
\begin{align}
P_{\txt{id}} 
&=  g \int \frac{d^3 \Pi }{(2\pi)^3}  ~T ~ \ln \Big[1 + e^{-\beta \big(\Pi^0 + U(n) - \mu\big)} \Big]~.
\end{align}
Using this form of $P$, it is straightforward to see that
\begin{align}
\bparr{P_{\txt{id}}}{\mu}_T &= g \int \frac{d^3 \Pi }{(2\pi)^3}  ~T  (-\beta) \bigg[ \bparr{U(n)}{\mu}_T - 1  \bigg] ~ \tilde{f} \nonumber \\
&= - n \bparr{U(n)}{\mu}_T + n~.
\label{P_id_derr}
\end{align}
Differentiation of the remaining term in Eq.\ \ref{eq:TD4} yields
\begin{align}
\parr{}{\mu}\bigg|_T \left(n U(n) - \int_0^{n} dn' ~  U(n') \right) &=  n \bparr{U(n)}{\mu}_T  ~,
\label{P_int_derr}
\end{align}
and, putting Eqs.\ \eqref{P_id_derr} and \eqref{P_int_derr} together, we confirm that $\left(\parr{P}{\mu}\right)_T = n$ is satisfied.

For a second test, we similarly check that $\inbparr{P}{T}_{\mu} = s$. First, we calculate
\begin{align}
\bparr{P_{\txt{id}}}{T}_{\mu} &= \frac{P_{\txt{id}}}{T} + g \int \frac{d^3 \Pi }{(2\pi)^3} ~ \frac{\Pi^0 + U(n) - \mu}{T} ~ \tilde{f} \nonumber \\
& \hspace{5mm} - ~ g \int \frac{d^3 \Pi }{(2\pi)^3}  ~ \bparr{U(n)}{T}_{\mu}   ~ \tilde{f} \nonumber \\
&= \frac{P_{\txt{id}}}{T} +  \frac{\mathcal{E}_{\txt{id}} + n U(n) - \mu n}{T}  \nonumber \\
& \hspace{5mm} - ~ n  \bparr{U(n)}{T}_{\mu}  ~.
\end{align}
A differentiation of the second term in the expression for the total pressure yields
\begin{align}
\parr{}{T}\bigg|_{\mu} \left(n U(n) - \int_0^{n} dn' ~  U(n') \right) &=  n \bparr{U(n)}{T}_{\mu} ~,
\end{align}
By putting the two above equations together and using Eq.\ \eqref{eq:effective_mu} we immediately obtain Eq.\ \eqref{eq:entropy_density} and complete the proof of the thermodynamic consistency of the model.

\section{Technical details of the simulations and tests of their influence on flow observables}
\label{sec:dependence_on_model_and_technical_details}

\subsection{Density calculation}
\label{sec:density_calculation}

The rest frame baryon density is obtained from $n_B^2 = (j_B)^{\mu} (j_B)_{\mu}$, with the baryon current $j_B^{\mu}$ computed according to
\begin{align}
j_B^{\mu}(\bm{r}) = \frac{1}{N_{\txt{ens}}} \frac{1}{N_T} 
 \sum_{\txt{ensembles}} \sum_{i=1}^{N_T} B_i
  \frac{p^{\mu}_i}{p^0_i} K_G(\bm{r} - \bm{r}_i, u_i) ~,
\end{align}
where $B_i$ is the baryon number of the $i$th species,
\begin{align}
K_G(\bm{r} - \bm{r}_i, u, \sigma) = \frac{u_0}{(2\pi \sigma^2)^{3/2}}
     \exp^{ -\frac{(\bm{r} - \bm{r}_i)^2 +
     \left(\bm{u} \cdot (\bm{r} - \bm{r}_i)\right)^2}{2\sigma^2}}
\end{align}
is a Lorentz-contracted Gaussian smearing kernel as described in \cite{Oliinychenko:2015lva}, with the width of the distribution set as $\sigma  = 1$ fm, and 
\begin{eqnarray}
u_i^{\mu} = (u_i^0, \bm{u}_i) = \frac{p_i^{\mu}}{m_i}
\end{eqnarray}
is a test-particle's four-velocity in the computational frame. The Gaussian smearing kernel is cut off at $2\sigma$ and renormalized to give $\int d^3r ~ K_G(\bm{r})  = 1$. Notice, however, that this does not guarantee that the sum over lattice nodes, $\sum_{r \in \text{lattice}} K_G(r)$, is equal to unity (this is because a sum over a finite number of values of $K_G(r)$ is in practice a numerical integration of $K_{G}(r)$ using a rectangle rule, which in the case of a Gaussian function introduces a numerical error); in consequence, baryon number is not conserved on the lattice exactly. We have also tested an alternative method of density calculation in which the smearing kernel in each Cartesian direction is a triangular function with the base proportional to the lattice spacing in that direction, often called ``triangular smearing'':
\begin{align}
K_t(\bm{r} - \bm{r}_i) = \frac{1}{(n^3 l_x l_y l_z)^2} g(\Delta x) g(\Delta y) g(\Delta z)~,
\end{align}
with 
\begin{eqnarray}
g(\Delta q) \equiv \big(n l_q - |\Delta q| \big) \theta\Big(n l_q - |\Delta q|\Big) ~,
\end{eqnarray}
where $l_x$, $l_y$, and $l_z$ are lattice spacings, and $n$ determines the range of smearing in units of lattice spacings; we used $n = 2$. The advantage of the triangular smearing is that $\sum_{r \in \text{lattice}} K_t(r)$ is equal to unity by construction (as integration by the rectangle rule is exact for a linear function), while the disadvantage is that it is not Lorentz covariant. Therefore, when triangular smearing is used, one needs to Lorentz-contract the lattice itself in the beam direction with the $\gamma$ factor corresponding to the beam energy (for simulations performed in the center-of-mass frame of the colliding nuclei). In the following, we will show that both types of smearing lead to comparable results provided that $N_{\txt{ens}}$ and $N_T$ are large enough to ensure a reliable density calculation; in this way, we show that our results do not depend on the technical implementation of the density calculation. 

We note here that both of our smearing paradigms (and, in fact, any smearing paradigm that we know of) lead to formal violations of causality. This is because the smearing kernels for calculating the density at a space-time point $(t,\bm{r})$ depend on the positions of the test particles in the vicinity of the point $\bm{r}$ at the same time $t$, implying instantaneous propagation of information about density changes within the smearing range.

For both types of smearing, the field derivatives $\partial^{\mu} A_{\nu}$ are computed numerically on the lattice, including the time derivative terms. The smearing kernels, mixed ensembles mode, and the equations of motion with the relativistic vector mean field $A^{\mu}(n_B)$ based on a polynomial form of $\alpha(n_B)$ (the VDF model) are included in the publicly available \texttt{SMASH} 2.1 software \cite{smash_version_2.1}; however, the parametrization of the vector mean field $A^{\mu}(n_B)$ with an arbitrary $c_s^2(n_B, T = 0)$ is not.

\subsection{Effects related to $N_T$ and $N_{\txt{ens}}$}
\label{sec:NT_effects}

\begin{figure}
    \centering
    \includegraphics[width=0.45\textwidth]{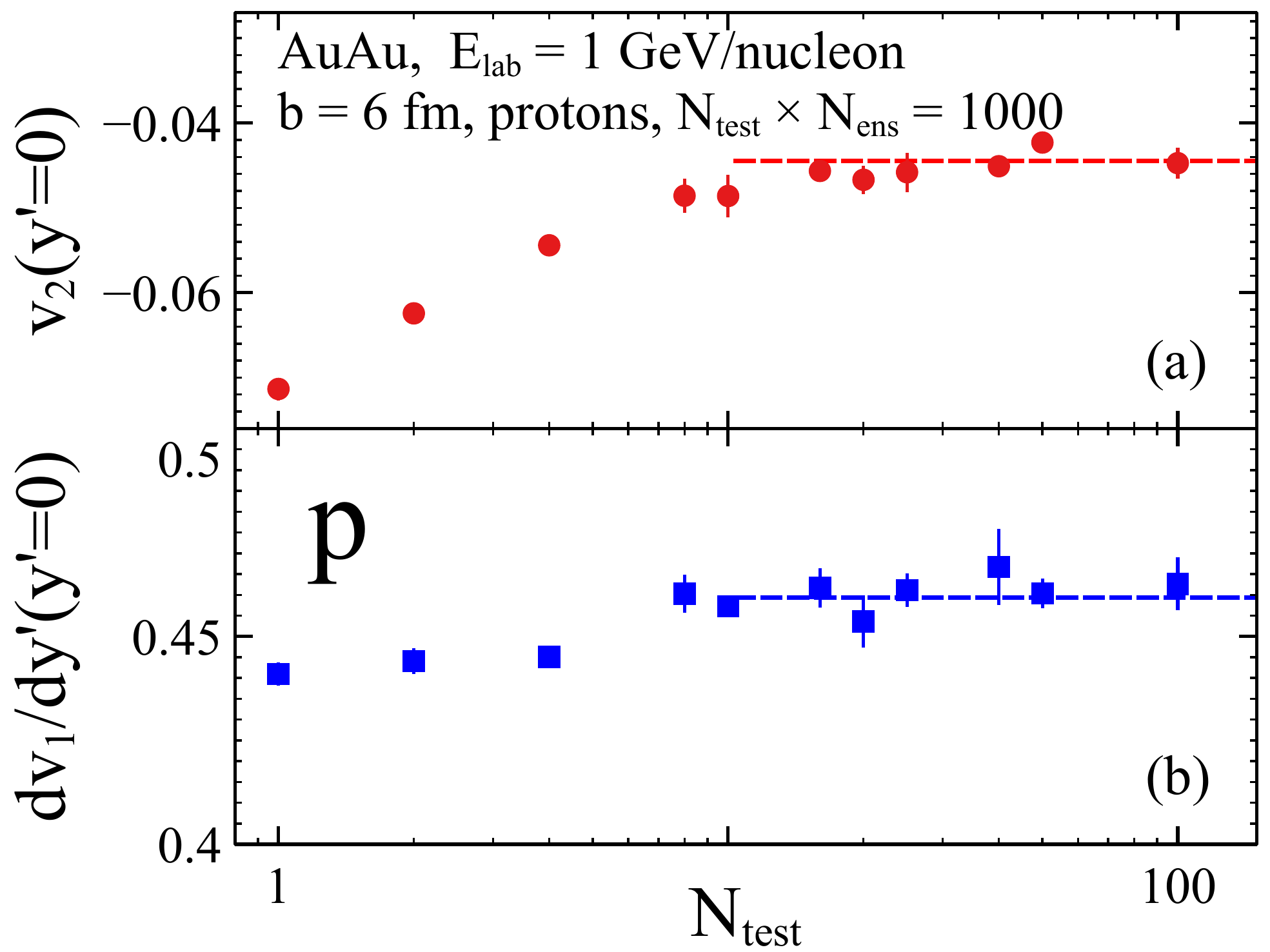}
    \caption{Effect of the number of test particles per particle $N_T$ on the proton elliptic flow (a) and the slope of the proton directed flow (b) in Au+Au collisions at $E_{\txt{lab}} = 1$ GeV/nucleon. The product $N_{T} N_{\txt{ens}} = 1000$ is kept constant, therefore the observed effect is primarily due to the dependence of collisions on $N_T$. Based on this result, for further simulations we choose $N_{T} = 20$.}
    \label{fig:ntest_p}
\end{figure}

\begin{figure}
    \centering
    \includegraphics[width=0.45\textwidth]{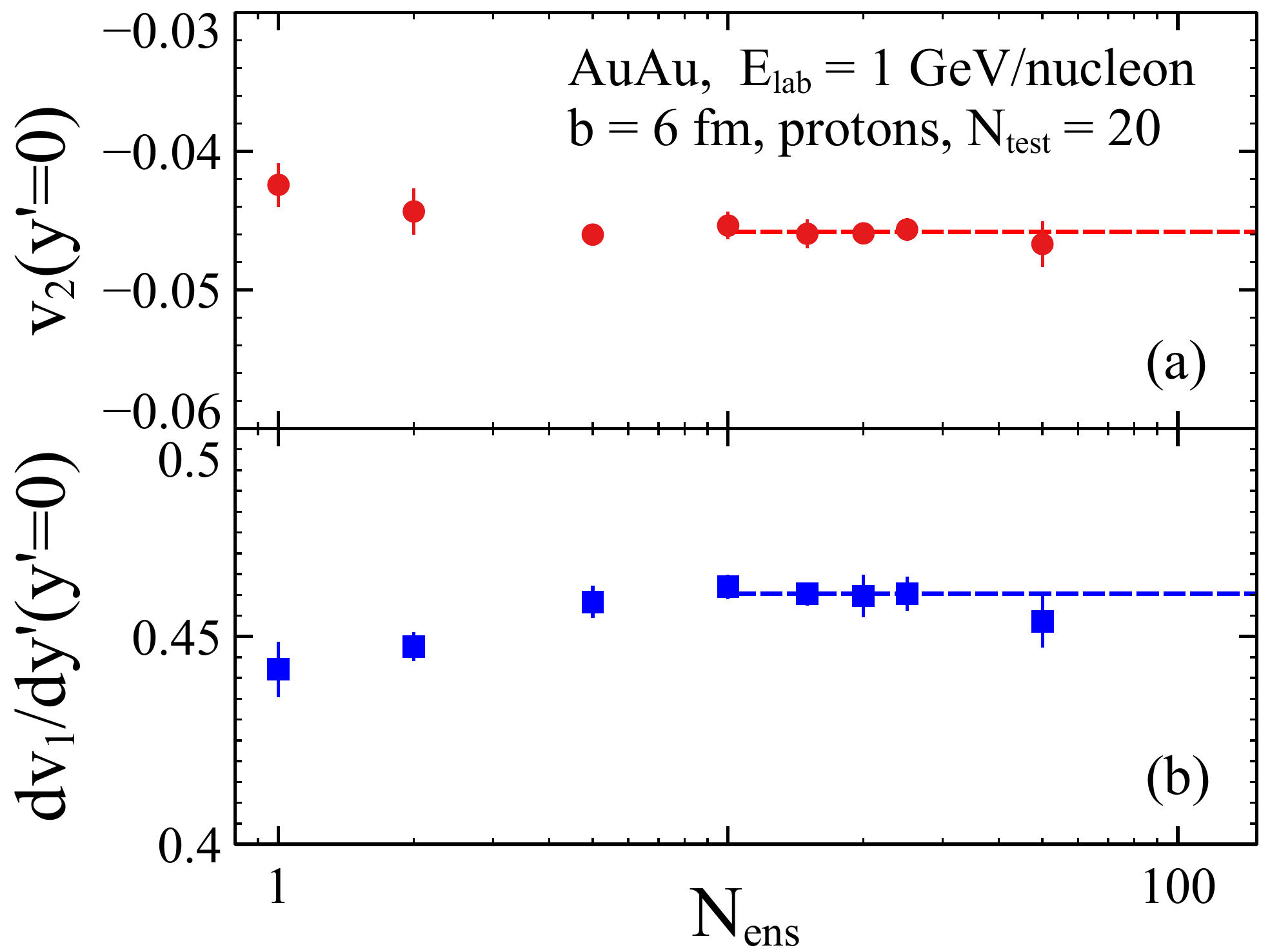}
    \caption{Effect of the number of ensembles $N_{\txt{ens}}$ on the proton elliptic flow (a) and the slope of the proton directed flow (b) in Au+Au collisions at $E_{\txt{lab}} = 1$ GeV/nucleon. The number of test particles per particle is set to $N_{T} = 20$, which, together with the  Gaussian smearing, seems to provide a rather accurate result even at $N_{\txt{ens}} = 1$ (see text for more details on the dependence of this result on the smearing paradigm). However, for better accuracy, we choose $N_{\txt{ens}} = 50$ for further simulations. }
    \label{fig:nens_p}
\end{figure}

We performed a comparison of results from simulations utilizing different numbers of test particles per particle $N_T$. To single out the effect of $N_T$ on collisions only and not on the accuracy of the density calculation (which we computed using the Gaussian smearing kernel), we kept the product $N_{T} N_{\txt{ens}}$ constant. Because the density and potentials are computed based on all test particles in all ensembles, by keeping $N_{T} N_{\txt{ens}} = \txt{const}$ we do not change the magnitude of the numerical spatial fluctuations in density, and thus we obtain the same accuracy of potential gradient calculations. In Fig.\ \ref{fig:ntest_p}, one can see that the effect of $N_{T}$ on proton flow observables is very substantial: as we show in the main text, the change of $v_2(y'=0)$ from -0.07 to -0.05 at $E_{\txt{lab}} = 1$ GeV/nucleon due to varying $N_T$ from 1 to 10 is comparable to a difference in $v_2(y'=0)$ obtained due to using a soft and a very stiff EOS. As the product $N_{T} N_{\txt{ens}}$ is kept constant in Fig.\ \ref{fig:ntest_p}, the observed effect of $N_{T}$ is entirely due to the dependence of collisions on $N_T$. The dependence of flow on $N_{T}$ is attributed to spurious effects of collisions at nonzero range \cite{Cheng:2001dz}, which become smaller at larger $N_{T}$ because cross sections scale with $N_T$ according to $\sigma \to \sigma / N_{T}$ and, therefore, the geometrical distance at which collisions occur decreases for larger $N_T$. In agreement with this interpretation, we observe a saturation of $v_2(y'=0)$ as a function of $N_{T}$ which occurs at $N_{T} \approx 15$ (while keeping $N_{T} N_{\txt{ens}} = 1000$). Our results show that any $N_{T} > 15$ is a reasonable choice, and we take $N_{T} = 20$ for our further simulations.

When $N_{T}$ is fixed, increasing the number of ensembles $N_{\txt{ens}}$ changes our results to a lesser extent, especially in the case of the elliptic flow, as shown in Fig.\ \ref{fig:nens_p}. This is likely because the Gaussian smearing together with the oversampling factor $N_T = 20$ provide an acceptably smooth spatial density profile even for $N_{\txt{ens}} = 1$, and the results saturate already at $N_{\txt{ens}} = 5$. We found that the computational cost increases almost proportionally to $N_{\txt{ens}}$ (specifically, it is slightly supralinear), while the simulation statistics increases exactly proportionally to $N_{\txt{ens}}$. Given the minimal additional computing cost, we chose to be on the safe side and we took $N_{\txt{ens}} = 50$. 

We note here that while similar tests of the influence of $N_T$ on flow observables have been performed within \texttt{SMASH} in the past (see, e.g., \cite{Mohs:2020awg}), the analysis presented here is qualitatively different as it separates the influence of scaled-down collision cross sections from the influence of the statistics used for calculating the mean field. In particular, as can be seen in Figs.\ \ref{fig:ntest_p} and \ref{fig:nens_p}, the two effects have an opposite influence on the magnitude of the elliptic flow in the explored collision energy region. In contrast, \cite{Mohs:2020awg} shows the effect of varying both the collision cross sections and the mean-field calculation statistics at the same time. Previous works exploring the influence of spurious fluctuations in the density calculation on the collision integral and the mean-field calculation include \cite{TMEP:2017mex,TMEP:2021ljz}; however, they do not address the effects on flow observables explicitly.

\begin{figure}
    \centering
    \includegraphics[width=0.45\textwidth]{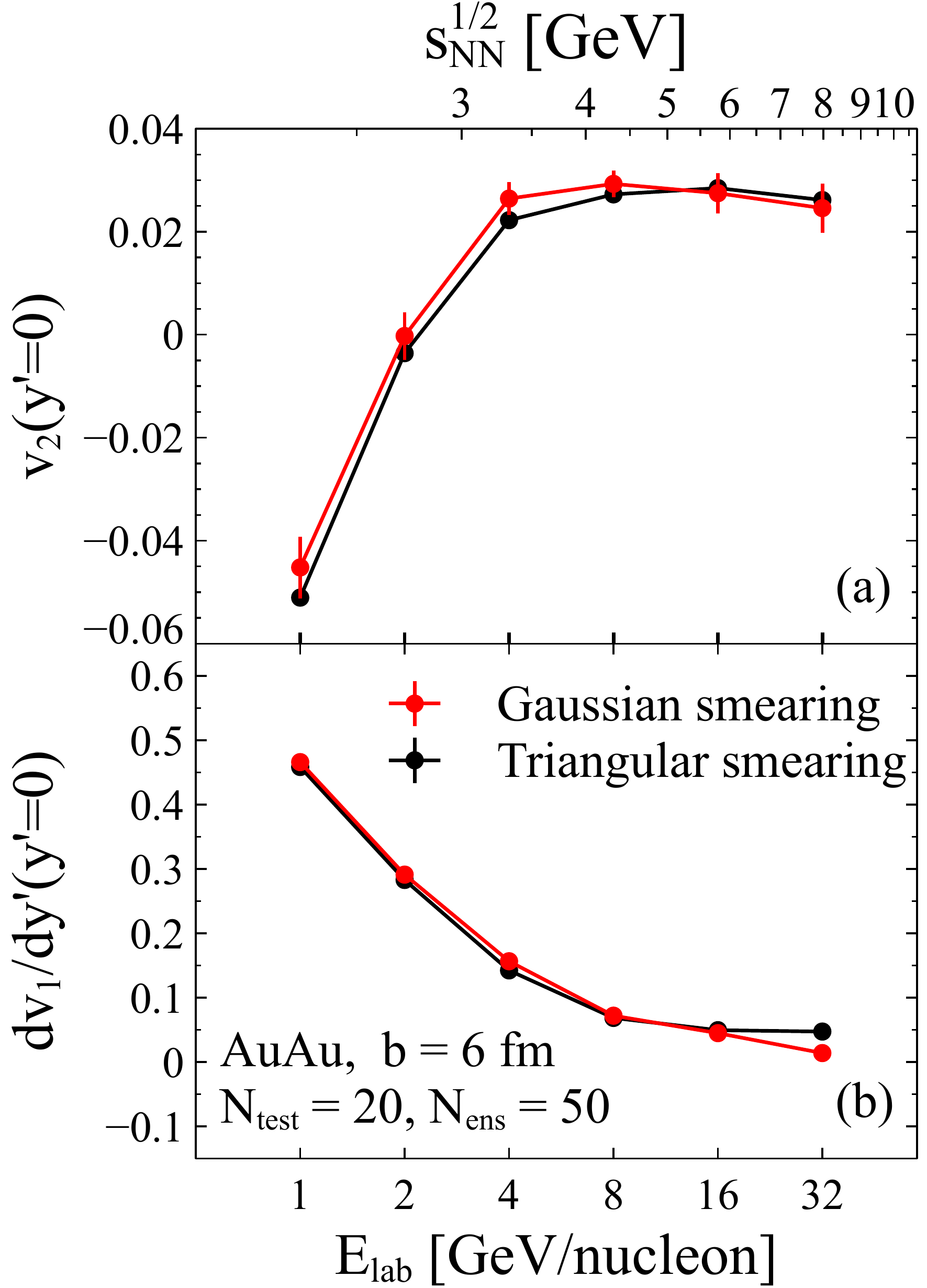}
    \caption{Comparison of the proton elliptic flow (a) and the slope of the proton directed flow (b) in Au+Au collisions as a function of the collision energy, obtained with simulations using different types of smearing kernels for computing baryon density on the Cartesian lattice.}
    \label{fig:smearing}
\end{figure}

We performed a similar test for simulations using the triangular smearing for the density calculation, and in this case the results suggested that this type of smearing demands a larger value of $N_{\txt{ens}}$; this is because the triangular smearing kernel is, by construction, less ``smooth'' than the Gaussian smearing kernel, and so it requires more statistics to yield a smooth density calculation. Nevertheless, in Fig.\ \ref{fig:smearing} we show that we obtain similar results for the proton flow using triangular and Gaussian smearing at $N_{T} = 20$ and $N_{\txt{ens}} = 50$. This is a non-trivial observation, because technically the two smearing paradigms are very different: in the Gaussian smearing the kernel is contracted along the direction of motion of a test particle, while in the triangular smearing the lattice is contracted along the beam direction. Naturally, since physical results should not be sensitive to the particular implementation of the density calculation, comparing the results obtained using two different smearing kernels is a quality test of our simulations. Finally, we note that the computing time is almost the same for simulations using the two smearing kernels, with a 4\% increase in the case of the triangular smearing; although the triangular smearing uses a kernel that is less computationally costly, in that case the lattice, which is more dense in the $z$ direction due to being contracted, needs a larger number of nodes in order to cover the same volume.

\subsection{Nucleus initialization}
\label{sec:nucleus_initialization}

Another important factor influencing the proton flow as obtained from the hadronic transport simulations is the initialization of the nuclei. We employ a default \texttt{SMASH} initialization as described in \cite{Weil:2016zrk}: in the coordinate space each nucleon is sampled independently from the Woods-Saxon distribution, while in the momentum space the nucleon momentum is sampled from a uniform distribution in the Fermi-sphere, where the radius of the Fermi sphere is computed from the local coordinate space density obtained analytically from the Woods-Saxon distribution. This simple model of nucleus initialization does not account for two physical phenomena: First, by construction, the mean field does not influence the initialization of a nucleus, and therefore the nucleus is not initialized exactly in a ground state. In freely moving nuclei, this results in ``breathing mode'' oscillations with the oscillation period on the order of 70--80 fm/$c$, as shown in Fig.\ 7 of \cite{Weil:2016zrk}. In simulations of heavy-ion collisions, the maximum compression is reached at the latest at 13 fm/$c$ (for $E_{\txt{lab}} = 1 ~ \txt{GeV/nucleon}$), as one can see in Fig.\ \ref{fig:I}, which is too short for the spurious oscillations to fully develop and thus does not lead to sizable contributions (especially given the fact that the differences in density due to the spurious oscillations are dwarfed by the changes in density due to the compression of the nuclei). Second, correlated nucleon pairs (as well as any other correlations) are not accounted for in the nucleus initialization, causing the high-momentum tail of the momentum distribution to be absent in our model, where the maximal momentum is limited by the Fermi momentum at the center of the nucleus. While we do not expect these phenomena to be important for the proton flow, we note that they may contribute to the overall systematic uncertainty of our simulations.

\subsection{Pauli blocking}
\label{sec:Pauli_Blocking}

In simulations of a dense baryonic medium one might expect Pauli blocking to influence the observables. However, even at $E_{\txt{lab}} = 1 ~ \txt{GeV/nucleon}$ the related effects turn out to be virtually absent. We find no significant difference in proton spectra or flow in simulations with and without Pauli blocking, which is in agreement with \cite{Weil:2016zrk}. The implementation of Pauli blocking in \texttt{SMASH} (described in \cite{Weil:2016zrk}) is based on estimating of the phase space density at the point of an action (where an action means a collision or a decay) and rejecting the action with probability $\prod_j (1 - f_j)$, where $f_j$ is the phase space density of particles $j$ exiting the action (a collision or a decay). We find that in Au+Au collisions at $E_{\txt{lab}} = 1 ~\txt{GeV/nucleon}$ with the impact parameter $b = 6$ fm, around 8\% of collisions and decays are blocked; at $E_{\txt{lab}} = 8 ~\txt{GeV/nucleon}$ this number is reduced to around 3\%.

\begin{figure}[t]
    \centering
    \includegraphics[width=0.45\textwidth]{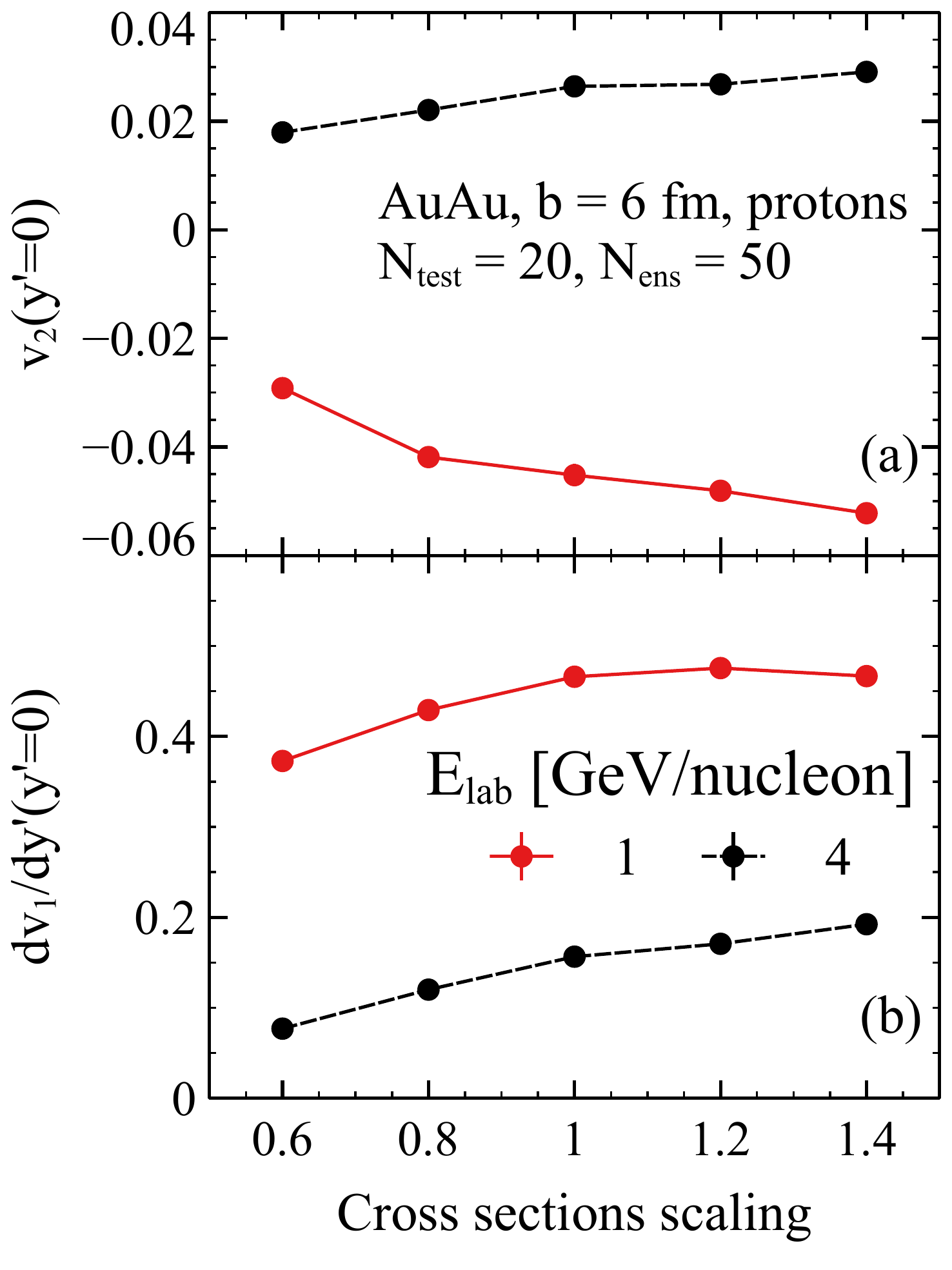}
    \caption{Sensitivity of the proton flow to scaling of all cross sections in Au+Au collisions at $b = 6$ fm, simulated using \texttt{SMASH}. At $n_B \in (0,2)n_0$, the mean-field is parametrized to reproduce the default Skyrme EOS, while at higher densities the mean field is parametrized to yield $c_s^2 \big[ n_B \in (2,3)n_0\big]= 0.1$ and $c_s^2 \big[ n_B >3n_0\big] = 0.3$.}
    \label{fig:xs_sensitivity}
\end{figure}

\subsection{Absence of isospin, Coulomb, and momentum-dependent interactions}
\label{sec:only_vector_potentials}

It is important to mention that the vector mean field $A^{\mu}$ is the only field implemented in our simulations. Coulomb interactions are not included, which may potentially influence our pion flow results \cite{HADES:2022mwn}. Isospin-dependent potentials likewise are not included, i.e., protons and neutrons feel the same mean field. At lower energies, in particular below $E_{\txt{lab}} = 0.5 ~\txt{GeV/nucleon}$, this cannot be justified, but at $E_{\txt{lab}} = 1 ~\txt{GeV/nucleon}$ and above it is a reasonable approximation due to the fact that matter at midrapidity is more isospin symmetric in that case and also because at high baryon density, the isospin-symmetric potential is much larger than the isospin-dependent one. A similar justification is valid for the absence of Coulomb potentials in our simulations.

It should also be mentioned that our vector mean field $A^{\mu}$ is a density-dependent field, but it includes no explicit dependence on the momentum of the particle that experiences it. This momentum dependence is necessary for the Schr\"{o}dinger-equivalent potential \cite{Jaminon:1981xg}, obtained from a given model, to reproduce the optical potential measured in elastic $p + A$ collisions \cite{Hama:1990vr}; also see \cite{Gale:1987zz,Weber:1992qc,Weber:1993et,Danielewicz:1999zn,Hong:2013yva} for discussion. While the measured optical potential exhibits growth with the energy of the particle, our equivalent optical potential, in fact, decreases. From a theoretical standpoint, the unphysical behavior of the equivalent optical potential is a substantial disadvantage of our model, especially in view of the effect of the momentum-dependent potentials on the $p_T$ dependence of $v_2$ shown in \cite{Danielewicz:1999zn} and the more recent analysis of the influence of various momentum-dependent potentials on flow observables \cite{Nara:2020ztb}. However, in this work we do not explore the momentum dependence of flow. Moreover, the integrated $v_2(y'=0)$ is not very sensitive to the momentum dependence of potentials in central and mid-central collisions, and only demonstrates a large sensitivity in peripheral collisions (see Fig.\ 4 of \cite{Danielewicz:1999zn}). Both the theory and the numerical implementation of a relativistic, explicitly momentum-dependent mean field are rather challenging, and while an example of such implementation is available in literature \cite{Weber:1992qc}, the adaptation of this approach to our case of flexible density-dependent potentials would still require a considerable theoretical effort.

\subsection{In-medium cross sections}
\label{sec:in-medium_cross_sections}

Some transport models \cite{Zhang:2010th,Gaitanos:2004ic} include density-dependent ``in-medium'' cross sections, that is, an \textit{ad hoc} modification of the vacuum cross sections reflecting an idea that some part of the interactions between particles is accounted for by the potentials and therefore the collision integral should be adjusted to reflect that fact. Here, we employ vacuum cross sections and do not make any in-medium modifications. However, we tested the effect of scaling of the cross sections on the flow observables, see Fig.\ \ref{fig:xs_sensitivity}. One can see that the difference in flow observables obtained by changing the scaling factor for all cross sections from $0.6$ to $1.4$ leads to rather moderate changes in both $dv_1/dy'(y'=0)$ and $v_2(y'=0)$. As one can see in Fig.\ \ref{fig:flow_sensitivity}, these changes are comparable to changing the speed of sound squared by 0.2, which one could take as the maximum systematic error in our estimate of the speed of sound squared due to variation of the cross sections.

\section{Symmetry energy expansion}
\label{sec:symmetry_energy_expansion}

In this appendix, we present a brief introduction to the role of the symmetry energy in the EOS of nuclear matter. For a review, see \cite{Baldo:2016jhp}.

The energy per particle in uniform nuclear matter can be decomposed into the following sum
\begin{align}
\vspace{-5mm}\frac{E}{N_B}(n_N, n_P) = \frac{E_0 }{N_B}(n_B) + S(n_B) \left(\frac{n_N - n_P}{n_B} \right)^2 + \dots~,
\label{eq:symmetry_energy_definition}
\end{align}
where $n_N$ and $n_P$ are the neutron and the proton density, respectively, such that $n_N + n_P = n_B$, $E_0 (n_B)$ is the energy of symmetric nuclear matter at $n_B$, and $S(n_B)$ is the (a)symmetry energy. Note that here, the energy is calculated with respect to the rest mass, such that $E/N = \infrac{\mathcal{E}}{n} - m_N$, where $\mathcal{E}$ is energy density that \textit{does} include the contribution from the rest mass. The symmetric part of the energy can be expanded around $n_B = n_0$ in the usual fashion,
\begin{align}
\frac{E_0}{N_B} (n_B) \approx E_{\txt{bin}} + \frac{K_0}{18}\left( \frac{n_B - n_0}{n_0} \right)^2 + \dots~,
\end{align}
where $E_{\txt{bin}} \approx  -16~ \txt{MeV}$ is the binding energy and $K_0 \equiv 9n_B^2 \big[\inderr{^2 \big( E/N_B \big)}{n_B^2}\big]\big|_{n_B = n_0}$ is the incompressibility (note that the linear term in the expansion disappears because $n_0$ is the equilibrium point). Similarly, one can expand $S(n_B)$ as
\begin{align}
S(n_B) \approx S_{0} + \frac{L}{3} \left( \frac{n_B - n_0}{n_0}\right) + \dots ~,
\label{eq:symmetry_energy_expansion}
\end{align}
where $S_{0}$ is the symmetry energy at $n_B = n_0$ and $L \equiv 3n_B \big[ \inderr{S}{n_B} \big]\big|_{n_B = n_0}$ is known as the slope of the symmetry energy at $n_B = n_0$.

By adding $m_N$ on both sides of Eq.\ \eqref{eq:symmetry_energy_definition}, multiplying by $n_B$, and inserting the symmetry energy expansion, Eq.\ \eqref{eq:symmetry_energy_expansion}, one arrives at
\begin{align}
\vspace{-5mm}\mathcal{E}= \mathcal{E}_0  +  \mathcal{E}_{\txt{sym}}~,
\end{align}
where $\mathcal{E}_0$ is the energy density of symmetric nuclear matter (including the kinetic energy of the system) and
\begin{align}
\mathcal{E}_{\txt{sym}}  =  n_B \left[ S_{0} + \frac{L}{3} \left( \frac{n_B - n_0}{n_0}\right) \right] \left(\frac{n_N - n_P}{n_B} \right)^2 + \dots ~.
\end{align}
Note that we can write
\begin{align}
\frac{n_N - n_P}{n_B} = 1 - 2Y_Q ~,
\end{align}
where $Y_Q = n_P / n_B$, so that 
\begin{align}
\mathcal{E}_{\txt{sym}}  =  n_B \left[ S_{0} + \frac{L}{3} \left( \frac{n_B - n_0}{n_0}\right) \right] \left(1 - 2Y_Q \right)^2 + \dots ~.
\end{align}

The pressure at $T=0$ is given by
\begin{align}
P = n_B^2 \derr{}{n_B} \bfrac{\mathcal{E}}{n_B} =  n_B \derr{\mathcal{E}}{n_B} - \mathcal{E} ~.
\end{align}
From the above equation it is evident that the pressure, like the energy density, can be written as a sum
\begin{align}
P = P_0 + P_{\txt{sym}} ~,
\end{align}
where $P_0$ is the part of the pressure coming from the symmetric part of the energy density (again, including the kinetic contribution) and $P_{\txt{sym}}$ is the asymmetric part, given by
\begin{align}
P_{\txt{sym}} = n_B \derr{\mathcal{E}_{\txt{sym}}}{n_B} - \mathcal{E}_{\txt{sym}} ~.
\end{align} 
Assuming that $Y_Q = \txt{const}$, one can immediately calculate
\begin{align}
P_{\txt{sym}} =   \frac{L}{3}  \frac{n_B^2 }{n_0} \big(1 - 2Y_Q \big)^2 ~.
\end{align} 

Overall, within the symmetry energy expansion, the (a)symmetry energy contributions can be added on top of the symmetric $\mathcal{E}_0$ and $P_0$. All other thermodynamic quantities can be then obtained in the standard way, including the speed of sound squared, $c_s^2\big|_{T=0} = \inbderr{P}{n_B}/\inbderr{\mathcal{E}}{n_B}$. The situation becomes especially simple in the case of pure neutron matter, for which $Y_Q = 0$.

\bibliography{inspire,noninspire}
\end{document}